\documentclass[10pt,journal]{IEEEtran}

\usepackage{cite}

\ifCLASSINFOpdf
\else
  \usepackage[dvips]{graphicx}
  \DeclareGraphicsExtensions{.eps}
\fi
\usepackage[cmex10]{amsmath}
\usepackage{amsfonts,amssymb,amsthm,mdwmath}
\newtheorem{lemma}{Lemma}

\usepackage{array}

\ifCLASSOPTIONcompsoc
  \usepackage[caption=false,font=normalsize,labelfont=sf,textfont=sf]{subfig}
\else
  \usepackage[caption=false,font=footnotesize]{subfig}
\fi

\usepackage{stfloats}
\usepackage{url}
\usepackage{color}
\usepackage{acronym}
\usepackage{enumitem}

\begin{document}
\acrodef{PPP}[PPP]{Poisson point process}
\acrodef{PGFL}[PGFL]{probability generating functional}
\acrodef{CDF}[CDF]{cumulative distribution function}
\acrodef{PDF}[PDF]{probability distribution function}
\acrodef{PMF}[PMF]{probability mass function}
\acrodef{PCF}[PCF]{pair correlation function}
\acrodef{RV}[RV]{random variable}
\acrodef{SIR}[SIR]{signal-to-interference ratio}
\acrodef{i.i.d.}[i.i.d.]{independent and identically distributed}
\acrodef{MRC}[MRC]{maximum ratio combining}
\acrodef{MAC}[MAC]{medium access control}
\acrodef{V2I}[V2I]{vehicular-to-infrastructure}
\acrodef{CoV}[CoV]{coefficient-of-variation}
\acrodef{1D}[1D]{one-dimensional}
\acrodef{LT}[LT]{Laplace transform}

\title{Performance of a Link in a Field of Vehicular Interferers with Hardcore Headway Distance}

\author{Konstantinos Koufos and Carl P. Dettmann 
\thanks{K.~Koufos and C.P.~Dettmann are with the School of Mathematics, University of Bristol, BS8 1UG, Bristol, UK. \{K.Koufos, Carl.Dettmann\}@bristol.ac.uk} \thanks{This work was supported by the EPSRC grant number EP/N002458/1 for the project Spatially Embedded Networks. All underlying data are provided in full within this paper.}}

\maketitle

\begin{abstract}
The Poisson point process (PPP) is not always a realistic model for the locations of vehicles along a road, because it does not account for the safety distance a driver maintains from the vehicle ahead. In this paper, we model the inter-vehicle distance equal to the sum of a constant hardcore distance and a random distance following the exponential distribution.  Unfortunately, the probability generating functional of this point process is unknown. To approximate the Laplace transform of interference at the origin, we devise simple approximations for the variance and skewness of interference, and we select suitable probability functions to model the interference distribution. When the coefficient-of-variation and the skewness of interference distribution are high, the PPP (of equal intensity) approximation of the outage probability becomes loose in the upper tail. Relevant scenarios are associated with urban microcells and highway macrocells with a low density of vehicles. The predictions of PPP deteriorate with a multi-antenna maximum ratio combining receiver and temporal indicators related to the performance of retransmission schemes. Our approximations generate good predictions in all considered cases. 
\end{abstract}


\section{Introduction}
Inter-vehicle communication, e.g., dedicated short-range transmission IEEE 802.11p, and/or connected vehicles to roadside units, e.g., LTE-based \ac{V2I} communication, will be critical for the coordination of road traffic, automated driving, and improved safety in emerging vehicular networks~\cite{V2X5G}. To analyze the performance of vehicular networks, we need tractable  but also realistic models for the locations of vehicles. The theory of point processes deals with random spatial patterns and can provide us with a general modeling framework~\cite{Veres2002}. 

With the advent of wireless communication networks with irregular structure~\cite{FCC2012}, elements from the theory of point processes have been employed to study their performance~\cite{Haenggi2009a}. The simplest model is a \ac{PPP} of some (potentially variable) intensity embedded in a mathematical space~\cite{Kingman1992}. The \ac{PPP} is characterized by complete randomness; the location of a point does not impose any constraints on the realization of the rest of the process. Due to the lack of inter-point interaction, the \ac{PGFL} is tractable~\cite[Theorem 4.9]{Haenggi2013a}, allowing us to calculate the probabilistic impact of suitable functions, e.g. an interference field, on the typical point. Because of that, the \ac{PPP} has been widely-adopted for performance evaluation, under interference, of heterogeneous wireless communication networks~\cite{Andrews2011, Dhillon2012, Haenggi2013b}.

Despite its wide acceptance, the \ac{PPP} has also received some criticism because it does not capture the repulsion between network elements due to physical constraints and \ac{MAC} mechanisms. This has sparked network modeling using many other point processes which, however, are not tailored to vehicular networks. For instance, stationary determinantal point processes fit better than \ac{PPP} the Ripley's K-function of macro-base station datasets~\cite{Andrews2015}. The \ac{PGFL} for the Ginibre and Gauss processes can be evaluated numerically~\cite{Andrews2015,Miyoshi2014,Deng2015}. The repulsion induced by collision avoidance \ac{MAC} is better captured by Mat{\'e}rn than softcore point processes~\cite{Busson2009, Haenggi2011}. The locations of users in wireless networks may also exhibit clustering due to non-uniform population density, and the \ac{PGFL} for some of the Poisson cluster processes is tractable, see~\cite[example 6.3(a)]{Veres2002} and~\cite{Ganti2009} for the \ac{PGFL} of the Neyman-Scott  process. 

A spatial model suitable for vehicular networks should be broken down into a model for the road infrastructure and another for the distribution of vehicles along each road. The Manhattan Poisson line process can model a regular layout of streets, while the Poisson line process is suitable for roads with random orientations. These processes have been coupled with homogeneous \ac{1D} \acp{PPP} for the distribution of vehicles along each line~\cite{Dhillon2018, Baccelli2018}. The \ac{LT} of interference for the line containing the typical receiver is calculated using the \ac{PGFL} of \ac{PPP}. The contribution of interference from other lines requires their distance distribution to the typical receiver~\cite{Dhillon2018}. Alternatively, one can map every line to the line containing the typical receiver with a non-uniform density of vehicles~\cite{Steinmetz2015}. The \ac{PPP} has also been used in \ac{1D} vehicular system setups for higher layer performance evaluation~\cite{Blaszczyszyn2013}. Non-homogeneous \acp{PPP} have been used to model the impact of mobility on temporal statistics of interference over finite regions~\cite{Koufos2016,Koufos2018a}.   

Unfortunately, the distribution of vehicles along a road with a few numbers of lanes, e.g., bidirectional traffic streams with restricted overtaking, will not resemble a \ac{PPP}. The \ac{PPP} allows unrealistically small headways with non-negligible  probability~\cite{Koufos2019} while in practice, the follower maintains a safety distance depending on its speed and reaction time plus the length of the vehicle ahead~\cite{Daou1966,Greenberg1966}. The distribution of headways naturally depends on traffic status. Measurements have revealed that the log-normal distribution is a good model under free-flow traffic, while the log-logistic distribution is adequate under congestion~\cite{Yin2009}. These distributions have been used to study the lifetime of inter-vehicle links~\cite{Yan2011}, however, without considering interference. 

In this paper, we would like to identify whether the \ac{PPP} for the locations of interfering vehicles adequately describes the performance of a link at the origin (not part of the point process generating the interference) and under which conditions. The simplest model that contains the \ac{PPP} as a special case, but can also be tuned to avoid small headways, consists of a constant hardcore distance plus a random component modeled by an exponential \ac{RV}~\cite{Cowan1975}. The inter-vehicle distance distribution becomes shifted-exponential, and the \ac{PPP} is obtained by setting the shift $c$ equal to zero. In~\cite[Fig.~8, Fig.~9]{Koufos2019}, we illustrated that the shifted-exponential distribution gives a good fit to real motorway traces, while the \ac{PPP} is a poor model near the origin (small headways). In~\cite{Koufos2018} we compared the variance and skewness of interference for the two models assuming equal intensity $\lambda$. We devised a simple formula that approximates the variance due to the shifted-exponential model being equal to the variance due to the \ac{PPP} multiplied by a factor that depends on $\lambda$ and $c$. Unfortunately, we could not relate the outage probabilities of the two models. Besides, the link performance over time, e.g., the mean local delay, and the outage probability with multi-antenna receivers require the temporal and spatial correlation properties of interference, which are different under the two deployment models. 

The common methodology to assess the average outage probability uses the \ac{PGFL} of the point process generating the interference. With a positive hardcore, the locations of vehicles become correlated, and we could not figure out how to use the \ac{PGFL} of \ac{PPP} as a building block for the calculation of the \ac{LT} of interference. Looking at the complications associated with the computation of the second and third moments of interference~\cite{Koufos2018}, it does not seem promising to calculate higher-order terms in the series expansion of \ac{PGFL}~\cite{Westcott1972}. The kernels in factorial moment representation of the \ac{PGFL} are simple only for the \ac{PPP}~\cite{Baccelli2012}. It has been recently shown that the outage probability in stationary wireless cellular networks can be well-approximated by horizontally shifting the outage probability due to a \ac{PPP}~\cite{Haenggi2015}. This is not applicable in our system setup because the point process does not impact the distribution of link gain but only the distribution of interference level. To assess the outage probability, we can also calculate a few moments of interference and select suitable distributions, with simple \ac{LT}, to approximate it. The method of moments has been widely used for modeling the signal-to-noise ratio in composite fading channels~\cite{Halim2010,Atapattu2011} and the aggregate interference in spectrum sensing~\cite{Sousa2008,Koufos2011}. 

In~\cite{Koufos2018}, we have shown that the \ac{CoV} of the interference level distribution (at the origin) is lower for the hardcore process~\cite{Koufos2018}. Complementing~\cite{Koufos2018}, we will generate a simple approximation for the skewness of interference too. Its sign would be crucial in selecting appropriate interference distribution models. These approximations would also allow us to deduce the traffic conditions where the \ac{PPP} fails to well-approximate the \ac{CoV} and the skewness of interference due to the hardcore process. Under these conditions, the \ac{PPP} will not accurately describe the interference distribution and the outage probability. To study the efficacy of \ac{PPP} with temporal and spatial performance metrics, we will respectively use the mean local delay and the outage probability of dual-branch \ac{MRC} receiver. The contributions of this paper are:
\begin{itemize}[leftmargin=*]
\item We approximate the distance distribution between the nearest interferer and the origin for a point process with hardcore distance $c$ and intensity $\lambda$. Its complexity rules out the possibility to calculate the signal level distribution for the $k$-th nearest interferer, and convert it to aggregate interference level by summing over all  $k\!\rightarrow\!\infty$.   
\item We show that for $\lambda c\!\ll\!1$, the skewness of interference is approximately equal to that due to a \ac{PPP} of intensity $\lambda$ scaled by $\left(1\!-\!\frac{\lambda c}{2}\right)$. Overall, a hardcore distance makes the distribution of interference more concentrated around the mean~\cite{Koufos2018}, and also less skewed. 
\item The skewness and the \ac{CoV} of interference increase for smaller cell sizes and lower intensity of vehicles. In these scenarios, the outage probability predicted using the \ac{PPP} is a poor approximation to the outage probability due to the hardcore process. On the other hand, the shifted-gamma distribution for the interference with parameters selected using the method of moments fits well the simulations.
\item Introducing hardcore distance reduces the spatial correlation of interference at the two branches of a \ac{MRC} receiver. Under \ac{i.i.d.} Rayleigh fading channels, the Pearson correlation coefficient scales down approximately by $\left(1\!-\!\lambda c\right)$. A bivariate gamma approximation for the interference distribution with identical and correlated marginals gives a good fit to the outage probability. 
\end{itemize}

The summary of the paper follows. In Section~\ref{sec:Model}, we present the system model. In Section~\ref{sec:Distance}, we derive the distance distribution to the nearest interferer. In Section~\ref{sec:Distribution}, we approximate the skewness of interference, and we select suitable \acp{PDF} for the distribution of interference. In Section~\ref{sec:Outage}, we illustrate that the approximations fit well the simulations, while the approximations using the \ac{PPP} may not be tight. In Section~\ref{sec:Applications}, we test the validity of the approximations using the mean local delay and a dual-antenna \ac{MRC} receiver. In Section~\ref{sec:Conclusions}, we conclude. 

\section{System model}
\label{sec:Model}
We consider \ac{1D} point process of vehicles $\Phi$, where the inter-vehicle distance follows the shifted-exponential \ac{PDF}. The shift is denoted by $c\!>\!0$, and the rate by $\mu\!>\! 0$. The intensity $\lambda$ of vehicles can be calculated from $\lambda^{-1}\!=\!c+\mu^{-1}$, or $\lambda\!=\!\frac{\mu}{1+\mu c}$. This model has been proposed by Cowan~\cite{Cowan1975}, and due to the positive shift $c$, it can avoid small inter-vehicle distances. The penalty paid is the correlations introduced in the locations of vehicles. The correlation properties have been studied in statistical mechanics, where the vehicles are the particles of \ac{1D} hardcore fluid, and the shift is equal to the diameter of the rigid disk modeling the particle~\cite{Salsburg1953}. The probability to find two particles at $x$ and $y\!>\!x$ is~\cite[equation~(32)]{Salsburg1953}:
\begin{equation}
\label{eq:rho2}
\rho_k^{\left(2\right)}\!\!\left(y,x\right) \!\!=\!\! \Bigg\{ \!\!\!\! \begin{array}{ccl} \lambda \! \sum\limits_{j=1}^k \!\! \frac{\mu^j \left(y-x-jc\right)^{j-1}}{\Gamma\!\left(j\right) e^{\mu\left(y-x-jc\right)}}, \!\!\!\!\!\!& &\!\!\!\!\!\! y\!\in\!\left(x\!\!+\!kc, x\!\!+\!\left(k\!+\!1\right)\!c\right) \\ 0, \!\!\!\!\!\!& &\!\!\!\!\!\! \text{otherwise}, \end{array} 
\end{equation}
where $k\!\geq\! 1$, $\Gamma\!\left(j\right)\!=\!\left(j\!-\!1\right)!$ and $\rho^{\left(2\right)}\!\!\left(y,x\right)\!=\!\sum\nolimits_{k=1}^\infty \rho_k^{\left(2\right)}\!\!\left(y,x\right)$. 
\begin{figure}[!t]
 \centering
  \includegraphics[width=\linewidth]{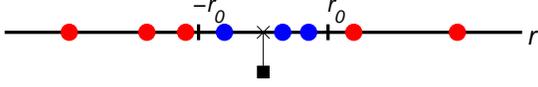} 
 \caption{The vehicles outside of the cell (red disks) generate interference at the receiver (black cross) located at the origin. The rest (blue disks) do not generate interference. A transmitter (black square) is paired with the receiver.}
 \label{fig:SystModel}
\end{figure}

The higher-order correlation functions are naturally more complicated. For a stationary determinantal point process, the $n$-th order correlation can be upper-bounded by the normalized $\left(n\!-\!1\right)$-th product of \acp{PCF} using Fan's inequality~\cite[Lemma 4]{Baccelli2012}. Fortunately, in \ac{1D} deployment, the inequality is tight~\cite[equation (27)]{Salsburg1953}. For instance, the third-order intensity measure describing the probability to find a triple of distinct vehicles at $x,y$ and $z$, is
\begin{equation}
\label{eq:rho3}
\rho^{\left(3\right)}\!\left(x,y,z\right) = \frac{1}{\lambda} \,  \rho^{\left(2\right)}\!\left(x,y\right) \rho^{\left(2\right)}\!\left(y,z\right), \, x\!<\!y\!<\!z.
\end{equation}

We place a receiver at the origin. The receiver and its associated transmitter are not part of the point process generating interference. The distance-based useful signal level $P_r$ is fixed and known. Only the vehicles outside a guard zone $\left[-r_0,r_0\right]$ contribute to interference, see Fig.~\ref{fig:SystModel}. For instance, in a \ac{V2I} communication, the vehicles inside the guard zone might be paired with the receiver (road side unit or traffic controller near an intersection), while the rest of the vehicles interfere with it. Furthermore, in a primary-secondary spectrum sharing scenario, the transmitter-receiver link might be a wireless backhaul using same spectral resources with vehicles communicating in ad hoc mode. The vehicles inside the guard zone are forced to stop their transmissions. Finally, calculating the interference at the origin with a guard zone around it would be useful in the modeling of other-lane interference due to beamforming transmissions~\cite[Fig.~13]{Koufos2019}. 

The transmit power level is normalized to unity. The propagation pathloss exponent is denoted by $\eta\!>\! 1$. The distance-based pathloss for an interferer located at $r$ is $g\!\left(r\right)\!=\!\left|r\right|^{-\eta}$ for $\left|r\right|\!>\!r_0$, and zero elsewhere. The fading power level over the interfering links, $h$, and over the transmitter-receiver link, $h_t$, is exponential (Rayleigh distribution for the fading amplitudes) with unit mean. Measurements have shown that the narrowband small scale fading in inter-vehicle communication resembles Rayleigh for distance separation in the order of $50-100$ m or more~\cite[Table III and IV]{Cheng2007}. The fading is \ac{i.i.d.} over different links and time slots. The interferers and the transmitter are active in each time slot, and they are equipped with single omni-directional antennas. When multiple antennas are employed at the receiver, they are separated at least by half the wavelength and their fading samples are \ac{i.i.d.} The distance-based pathlosses to the two antennas are assumed equal. 

\section{Distance distributions}
\label{sec:Distance}
The statistics of interference at the origin are closely related to the statistics of the distance to the interferers. Let us denote by $X_1$ the \ac{RV} describing the distance to the nearest interferer. For the \ac{PPP}, due to the independence property, it suffices to calculate the distance distribution without the guard zone and shift it by $r_0$. The contact distribution for a \ac{1D} \ac{PPP} is exponential with parameter twice the intensity, $f_{X_1}\!\left(x\right)\!=\! 2\lambda e^{-2\lambda \left(x-r_0\right)}, x\!\geq\!r_0$. It is straightforward to verify that the \ac{CoV} and the skewness of $X_1$ for the \ac{PPP} are equal to $\frac{1}{1+2\lambda r_0}$ and two respectively. 

For the hardcore process, the distribution of $X_1$ follows easily, if we ignore the guard zone. For $x\!\leq\!\frac{c}{2}$, the point process rules out any other interferer closer than $x$ to the origin. The nearest interferer is located uniformly in $\left[-\frac{c}{2},\frac{c}{2}\right]$. The probability to find a vehicle within an infinitesimal ${\rm d}x$ is $\lambda {\rm d}x$. As a result, the distance distribution is uniform in $\left[0,\frac{c}{2}\right]$, and the probability to observe any distance of this range is $2\lambda {\rm d}x$. Therefore $\mathbb{P}\!\left(X_1\!\leq\! \frac{c}{2}\right)\!=\!\lambda c$. For $x\!\geq\!\frac{c}{2}$, no other interferer must be located within a distance $\left(2x\!-\!c\right)$, thus  $\mathbb{P}\!\left(X_1\!\geq\! x|x\!\geq\!\frac{c}{2}\right)\!=\!e^{-\mu\left(2x-c\right)}$. After deconditioning, $\mathbb{P}\!\left(X_1\!\geq\! x\right)\!=\!\left(1\!-\!\lambda c\right)e^{-\mu\left(2x-c\right)}, x\!\geq\!\frac{c}{2}$. Finally, the \ac{CDF} for the \ac{RV} $X_1$ becomes 
\[
\mathbb{P}\!\left(X_1\!\leq\! x\right) = \Big\{ \begin{array}{ccl} 2\lambda x, \!\!\!\!\!& &\!\!\!\!\! x\!\in\! \left[0,\frac{c}{2}\right) \\  1\!-\!\left(1\!-\!\lambda c\right) e^{-\mu\left(2x-c\right)}, \!\!\!\!\!& &\!\!\!\!\! x\!\geq\!\frac{c}{2}.  \end{array}
\]

The guard zone raises the complexity of calculating the distribution of $X_1$ because the locations of vehicles from the two sides of the guard zone are correlated. However, in a practical system setup, this correlation should be weak. In order to give a relevant approximation, we note that a high value for the dimensionless ratio $\frac{\mu}{\lambda}\!=\!\frac{1}{1-\lambda c}\!=\left(1\!+\!\mu c\right)\!\gg\!1$ indicates that the \ac{PCF} decorrelates slowly. The point process decorrelates within a distance $\frac{2 r_0}{c}$, if $\frac{2 r_0}{c}\!\gg\!\frac{\mu}{\lambda}\!=\!\left(1\!+\!\mu c\right)\!\approx\! \mu c$, or equivalently, $\mu\!\ll\!\frac{2r_0}{c^2}$. If this condition is true, we introduce minor error by treating as \ac{i.i.d.} the distances of the nearest interferer from opposite sides of the guard zone. 

Let us denote by $X_1^{\text{p}}$ the \ac{RV} describing the distance of the nearest interferer from the positive half-axis. For $x\!\in\!\left(r_0,r_0\!+\!c\right)$, $X_1^{\text{p}}$ follows the uniform distribution. For $x\!\geq\!r_0\!+\!c$, no other interferer must be located closer to the cell border, $\mathbb{P}\!\left(X_1^{\text{p}}\!\geq\! x|x\!\geq\!r_0\!+\!c\right)\!=\!e^{-\mu\left(x-r_0-c\right)}$, or $\mathbb{P}\!\left(X_1^{\text{p}}\!\geq\! x\right)\!=\!\left(1\!-\!\lambda c\right)e^{-\mu\left(x-r_0-c\right)}, x\!\geq\!r_0\!+\!c$ after deconditioning.  Finally, 
\[
\mathbb{P}\!\left(X_1^{\text{p}}\!\leq\! x\right) = \Big\{ \begin{array}{ccl} \lambda\left(x\!-\!r_0\right), \!\!\!\!\!& &\!\!\!\!\! x\!\in\! \left[r_0,r_0\!+\!c\right) \\  1\!-\!\left(1\!-\!\lambda c\right) e^{-\mu\left(x-r_0-c\right)}, \!\!\!\!\!& &\!\!\!\!\! x\!\geq\!r_0\!+\!c.  \end{array}
\]

The approximation for the \ac{CDF} of $X_1$ follows from the distribution of the minimum of two \ac{i.i.d.} \acp{RV} $X_1^{\text{p}}$. 
\begin{equation}
\label{eq:DistanceCDF}
\mathbb{P}\!\left(\!X_1\!\leq\! x\!\right) \!\approx\! \bigg\{\!\! \begin{array}{ccl} \!\! 1\!-\!\left(\!1\!-\lambda\left(x\!-\!r_0\right)\right)^2, \!\!\!\!\!& &\!\!\!\!\! x\!\in\! \left[r_0,r_0\!+\!c\right) \\  \!\! 1\!-\!\left(1\!-\!\lambda c\right)^2 \!\! e^{-2\mu\left(x-r_0-c\right)}, \!\!\!\!\!& &\!\!\!\!\! x\!\geq\!r_0\!+\!c.  \end{array}
\end{equation}
\begin{figure*}[!t]
 \centering \subfloat[]{\includegraphics[width=\columnwidth]{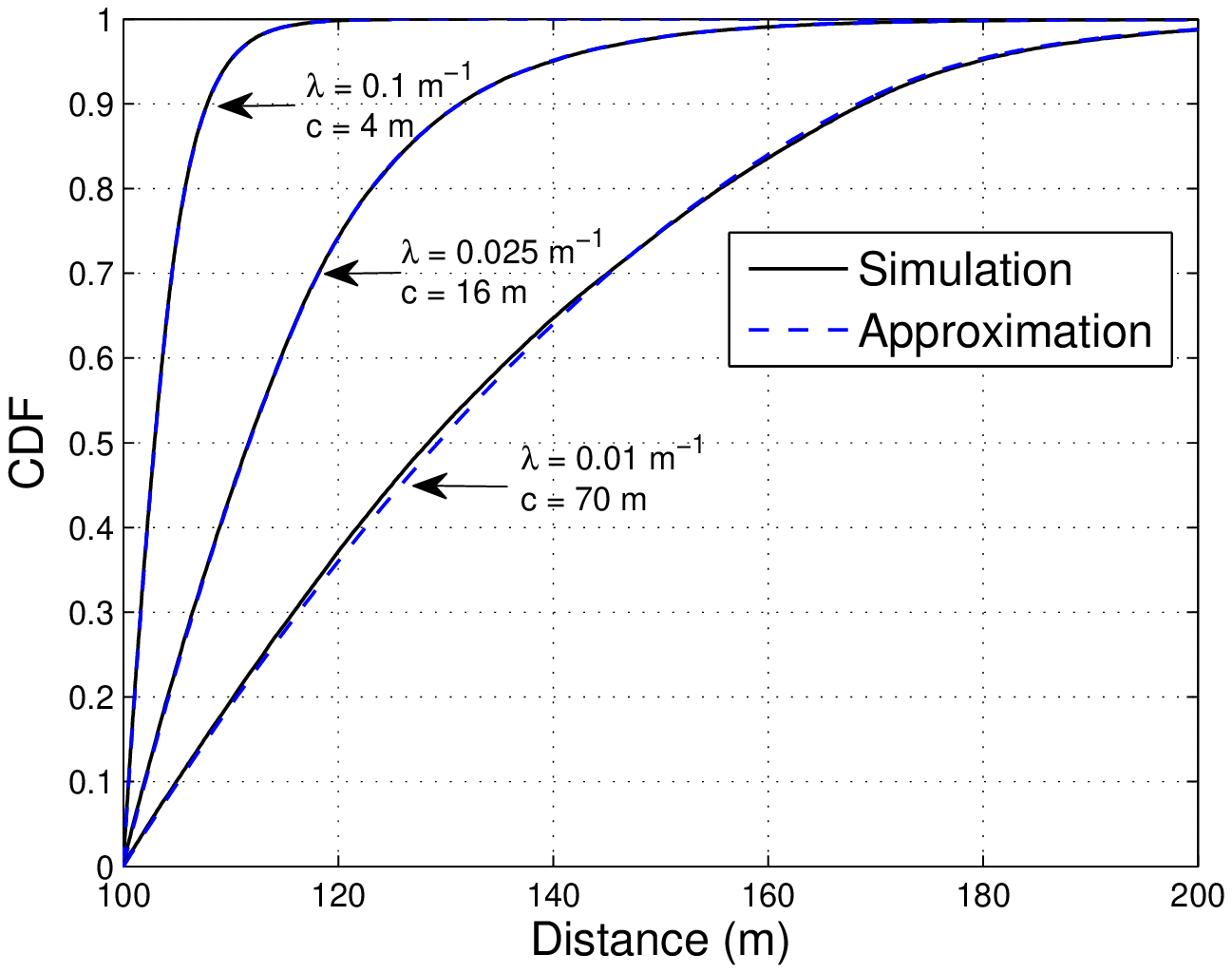}\label{fig:NearestDistanceA}}\hfil \subfloat[]{\includegraphics[width=\columnwidth]{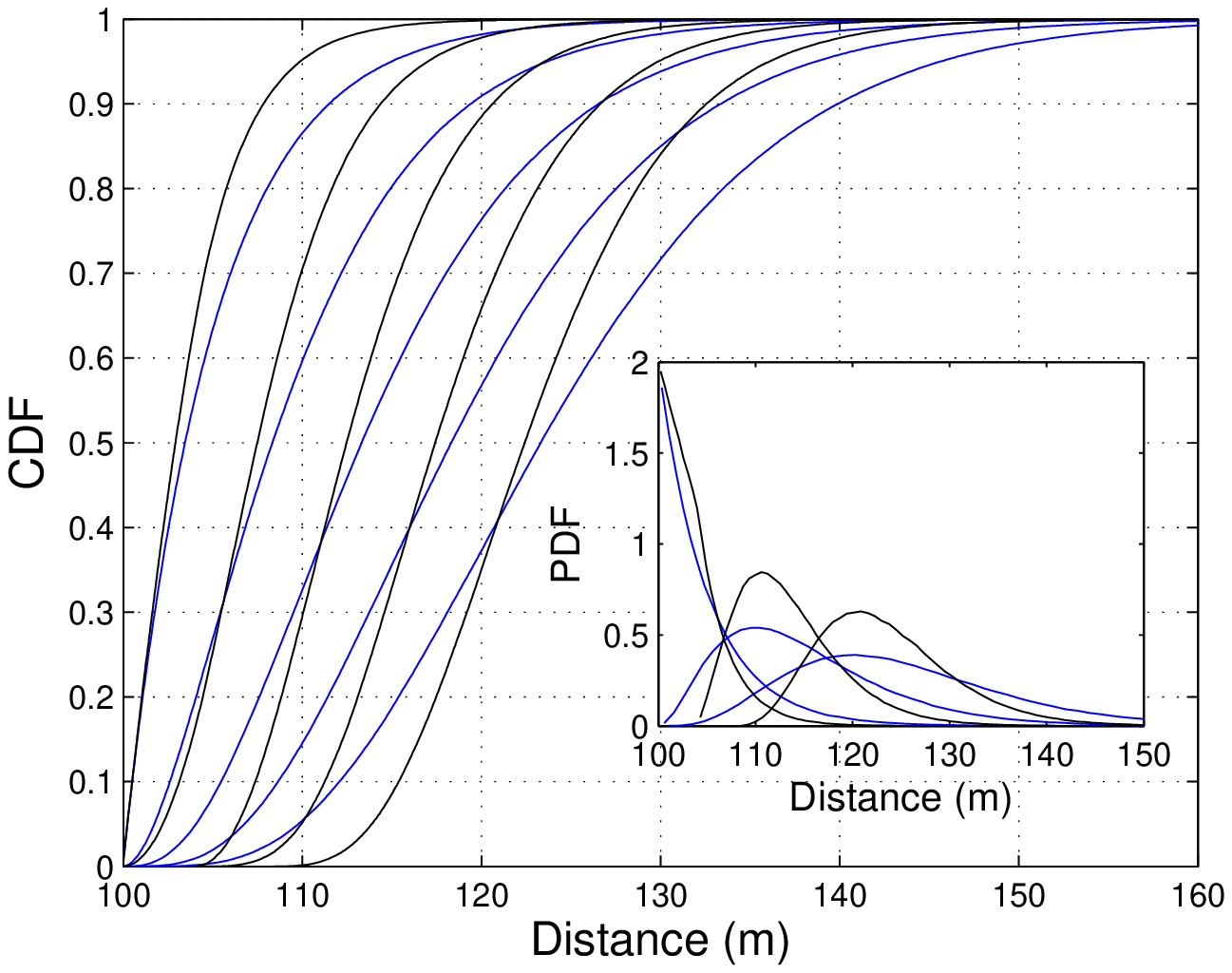}\label{fig:NearestDistanceB}} 
 \caption{(a) The \ac{CDF} of the distance between the nearest interferer and the origin. The approximation in~\eqref{eq:DistanceCDF} is verified against the simulations for $\lambda c\!=\!0.4$, $\lambda c\!=\!0.7$ and $r_0\!=\! 100$ m. (b) Simulated \ac{CDF} of the distance between the $k-$th nearest interferer and the origin for a \ac{PPP} of intensity $\lambda\!=\!0.1$ (blue lines) and for a hardcore process with $\lambda\!=\! 0.1\,{\text{m}}^{-1}$ and $c\!=\! 4\, {\text{m}}$ (black lines). Guard zone $r_0\!=\! 100$ m. In the inset, the associated \acp{PDF} are depicted for $k\!\in\!\left\{1,3,5\right\}$.}
 \label{fig:NearestDistance}
\end{figure*}

The above approximation overlaps with the simulations even if the mean inter-vehicle distance, $\lambda^{-1}\!=\! 40$ m, becomes comparable to the guard zone, $r_0\!=\!100$ m, see Fig.~\ref{fig:NearestDistanceA}. This is because for $\lambda c\!=\! 0.4$, the \ac{PCF} converges to $\lambda^2$ after approximately $4c$, see~\cite[Fig.~3]{Koufos2018}, which is equal to $64$ m, roughly one-third of the guard zone length. Even for the unrealistically high value of the product $\lambda c\!=\! 0.7$, along with a large hardcore $c\!=\!70\, {\text{m}}$, the error is negligible.  

\noindent 
Differentiating~\eqref{eq:DistanceCDF}, the approximation for the \ac{PDF} becomes 
\begin{equation}
\label{eq:DistancePDF}
f_{X_1}\!\!\left(x\right) \approx \Big\{ \begin{array}{ccl} 2\lambda\left(\!1\!-\lambda\left(x\!-\!r_0\right)\right), \!\!\!\!\!& &\!\!\!\!\! x\!\in\! \left[r_0,r_0\!+\!c\right) \\  2\lambda\left(1\!-\!\lambda c\right) e^{-2\mu\left(x-r_0-c\right)}, \!\!\!\!\!& &\!\!\!\!\! x\!\geq\!r_0\!+\!c.  \end{array}
\end{equation}

It is possible to verify that the \ac{CoV} and the skewness of~\eqref{eq:DistancePDF} are less than $\frac{1}{1+2\lambda r_0}$ and two respectively, the values associated with a \ac{PPP} of equal intensity. In Fig.~\ref{fig:NearestDistanceB} we have simulated the distance distribution for the $k-$th nearest interferer, $k\!\leq\! 5$, which follows the same trend as that proved for $k\!=\!1$: The distributions of the  hardcore process have lower \ac{CoV} and skewness as compared to those of \ac{PPP}. This complies with the intuition that a hardcore makes the point process less random. Based on this, we may conjecture that the distribution of interference due to the hardcore process will be more concentrated around the mean and also less skewed as compared to that due to a \ac{PPP} of equal intensity. 

\section{Interference distribution}
\label{sec:Distribution}
Using Campbell's Theorem, we can calculate the mean interference at the origin due to a stationary point processs of intensity $\lambda$ within $\left(-\infty,-r_0\right)\!\cup\!\left(r_0,\infty\right)$: $\mathbb{E}\!\left\{\mathcal{I}\right\} \!=\!2\lambda\int_{r_0}^\infty \!x^{-\eta}{\rm d}x \!=\! \frac{2\lambda r_0^{1-\eta}}{\eta-1}$. The details for the approximation of the second moment can be found in~\cite[Section V]{Koufos2018}. Over there we have approximated the \ac{PCF} with that due to a \ac{PPP} for large  distance separation, $\rho^{(2)}\!\left(x,y\right) \!\approx\!\lambda^2, \left|y\!-\!x\right|\!>\!3c$, and used the exact \ac{PCF} for smaller distances. This approximation should be valid for $\lambda c\!\ll\! 1$~\cite[Fig.~3]{Koufos2018}. The hardcore process is less random than the \ac{PPP}, and the variance of inteference reduces to~\cite[equation (14)]{Koufos2018}:  
\begin{equation}
\label{eq:VarApp}
\begin{array}{ccl}
\mathbb{V}\!\left\{\mathcal{I}\right\} \!\!\!&\approx&\!\!\! \displaystyle \frac{4\lambda r_0^{1-2\eta}}{2\eta\!-\!1} \left(1\!-\!\lambda c\!+\!\frac{1}{2}\lambda^2c^2\right), 
\end{array}
\end{equation}
where the term in front of the parenthesis is the variance due to a \ac{PPP} of intensity $\lambda$. 

Some preliminary calculations of the third moment are available in~\cite[Section IV]{Koufos2018}. Next, we derive a simple approximation relating it to that due to a \ac{PPP} of equal intensity. 
\begin{lemma}
\label{lemma:1}
The skewness of interference from a hardcore process of intensity $\lambda$ and hardcore distance $c$ can be approximated by the skewness due to a \ac{PPP} of intensity $\lambda$, scaled by $\left(1\!-\!\frac{\lambda c}{2}\right)$. The approximation is valid for $\lambda c\!\rightarrow\!0$ and $\frac{c}{r_0}\!\rightarrow\!0$. 
\[
\mathbb{S}\!\left\{\mathcal{I}\right\} \approx \frac{12 \lambda r_0^{1-3\eta}}{3\eta\!-\!1} \left( \frac{4\lambda r_0^{1-2\eta}}{2\eta\!-\!1} \right)^{-\frac{3}{2}} \left(1 \!-\! \frac{\lambda c}{2} \right).
\]
\begin{proof}
The proof can be found in the appendix. 
\end{proof}
\end{lemma}

Few properties can be drawn from Lemma~\ref{lemma:1}: (i) Introducing hardcore distance while keeping the intensity of interferers fixed reduces the skewness but the interference distribution remains positively-skewed. (ii) The skewness of interference due to a \ac{PPP} increases for increasing pathloss exponent $\eta$ and decreasing cell size $r_0$. Introducing hardcore distance for fixed $\lambda$ does not change this property. (iii) Increasing the intensity $\lambda$ reduces the skewness of interference for fixed $c$. 
\begin{figure}[!t]
 \centering
  \includegraphics[width=\linewidth]{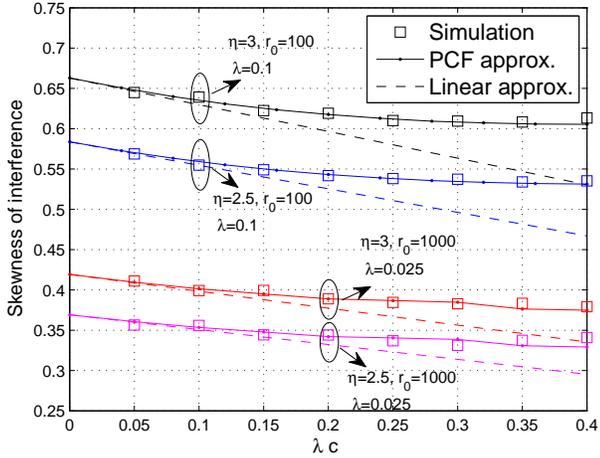}
 \caption{Skewness with respect to $\lambda c$ for urban microcells, $r_0\!=\!100$ m, and motorway macrocells, $r_0\!=\! 1$ km. In macrocells, we expect higher speeds, thereby lower intensity $\lambda$ and larger tracking distances $c$. For the \ac{PCF} approximations we used $\rho^{(2)}\!\left(x,y\right) \!\approx\!\lambda^2, \left|y\!-\!x\right|\!>\!2c$. The linear approximation refers to the result in Lemma~\ref{lemma:1}.}
 \label{fig:SkewnProp}
\end{figure}

The above properties can be observed in Fig.~\ref{fig:SkewnProp}, where we have simulated the skewness for different cell size $r_0$, pathloss exponent $\eta$ and traffic parameters $\lambda, c$. We see that for the considered range of $\lambda c$, the approximations for the second- and the third-order correlation, $\rho^{\left(2\right)}\!\left(x,y\right), \rho^{\left(3\right)}\!\left(x,y,z\right)$ do not introduce practically any error as compared to the simulations. In addition, the approximation given in Lemma~\ref{lemma:1} is quite accurate for small $\lambda c$. While changing from the microcell to macrocell scenario, we have the interplay of two conflicting factors: On one hand, the intensity of vehicles decreases to account for the higher speed of vehicles, and this increases the skewness. On the other hand, the cell size increases which reduces the skewness. According to Lemma~\ref{lemma:1}, the skewness is proportional to $\frac{1}{\sqrt{\lambda r_0}}$. For the selected parameter values of Fig.~\ref{fig:SkewnProp} the skewness is smaller for the macrocells. 

For a bounded pathloss model, the interference distribution strongly depends on the fading process~\cite{Kountouris2014}. In our system setup we note: (i) the positive skewness of interference, and (ii) the guard zone around the receiver (which essentially bounds the pathloss model) along with the exponential \ac{PDF} for the fading. The gamma \ac{PDF}  has a positive skewness, and it includes the exponential \ac{PDF} as a special case. The parameters $k,\beta$,  $f_{\mathcal{I}}\!\left(x\right) \approx \frac{x^{k-1} e^{-x/\beta}} {\Gamma\left(k\right) \beta^k}$, can be computed by matching the mean and the variance approximation in~\eqref{eq:VarApp}, resulting to $k\!=\!\frac{\mathbb{E}\left\{\mathcal{I}\right\}^2} {\mathbb{V}\left\{\mathcal{I}\right\}}$ and $\beta\!=\!\frac{\mathbb{E}\left\{\mathcal{I}\right\}}{k}$. The skewness of the gamma distribution is $\frac{2}{\sqrt{k}}$. For practical values of the pathloss exponent $\eta\!\in\!\left[2,6\right]$ and realistic traffic parameters $\lambda c\!<\!\frac{1}{2}$~\cite[Fig.~8, Fig.~9]{Koufos2019}, one can verify that the skewness, $\frac{2}{\sqrt{k}}$, is less than the approximation given in Lemma~\ref{lemma:1}. The shifted-gamma \ac{PDF}, which matches also the skewness of interference, is expected to provide better fit than the gamma \ac{PDF}. 
\[
f_{\mathcal{I}}\!\left(x\right) \approx \frac{\left(x\!-\!\epsilon\right)^{k-1} e^{-\left(x-\epsilon\right)/\beta}} {\Gamma\left(k\right) \beta^k}, \, x\!\geq\! \epsilon, 
\]
where $k\!=\!\frac{4}{\mathbb{S}\left\{\mathcal{I}\right\}^2}, \beta\!=\!\sqrt{\frac{\mathbb{V}\left\{\mathcal{I}\right\}}{k}}$ and shift $\epsilon\!=\! \mathbb{E}\!\left\{\mathcal{I}\right\}\!-\!k\beta$. 

\begin{figure*}[!t]
 \centering \subfloat[$\lambda=0.1\, {\text{m}}^{-1}, c\!=\! 4\, {\text{m}}$]{\includegraphics[width=\columnwidth]{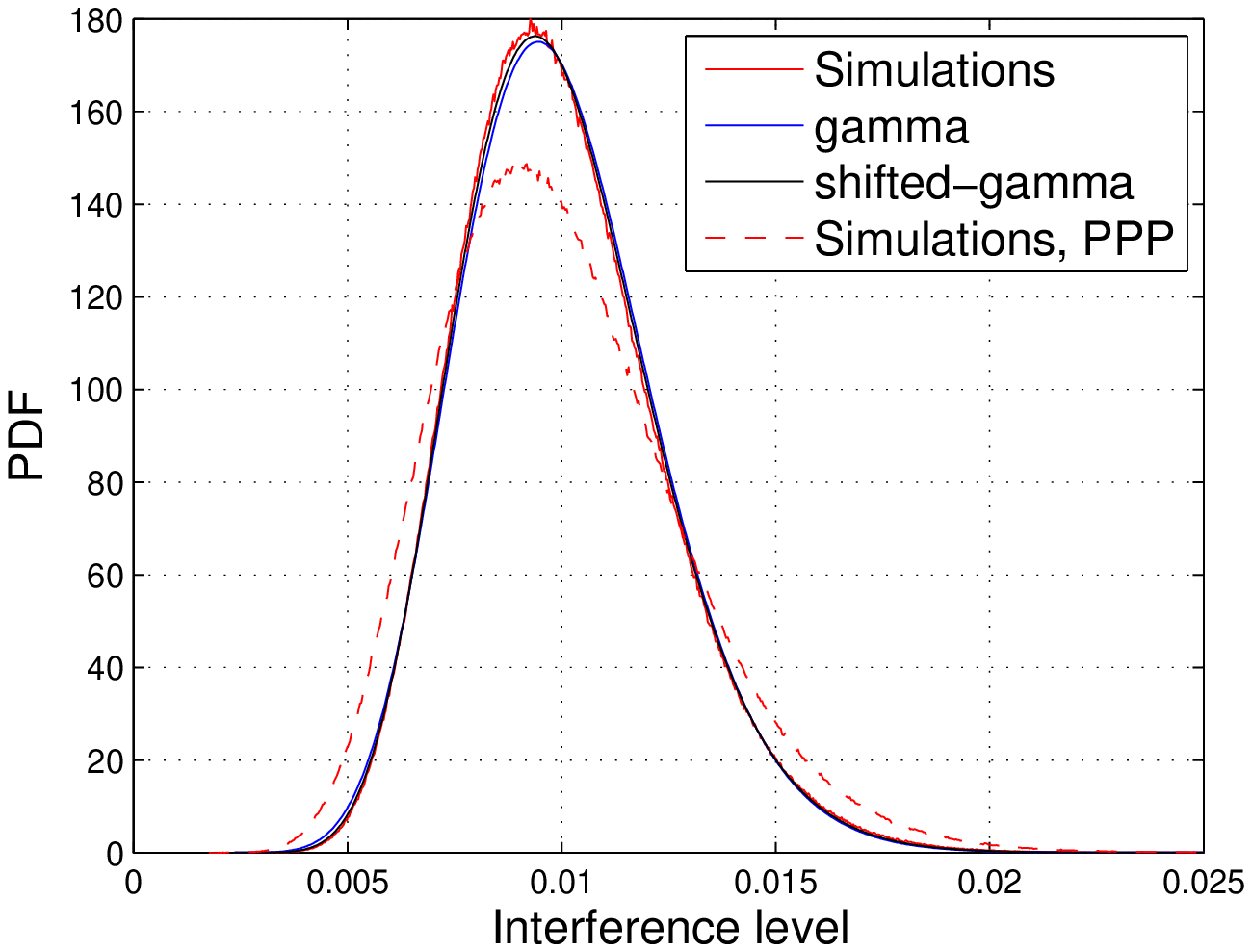}}\hfil \subfloat[$\lambda=0.025\, {\text{m}}^{-1}, c\!=\! 16\, {\text{m}}$]{\includegraphics[width=\columnwidth]{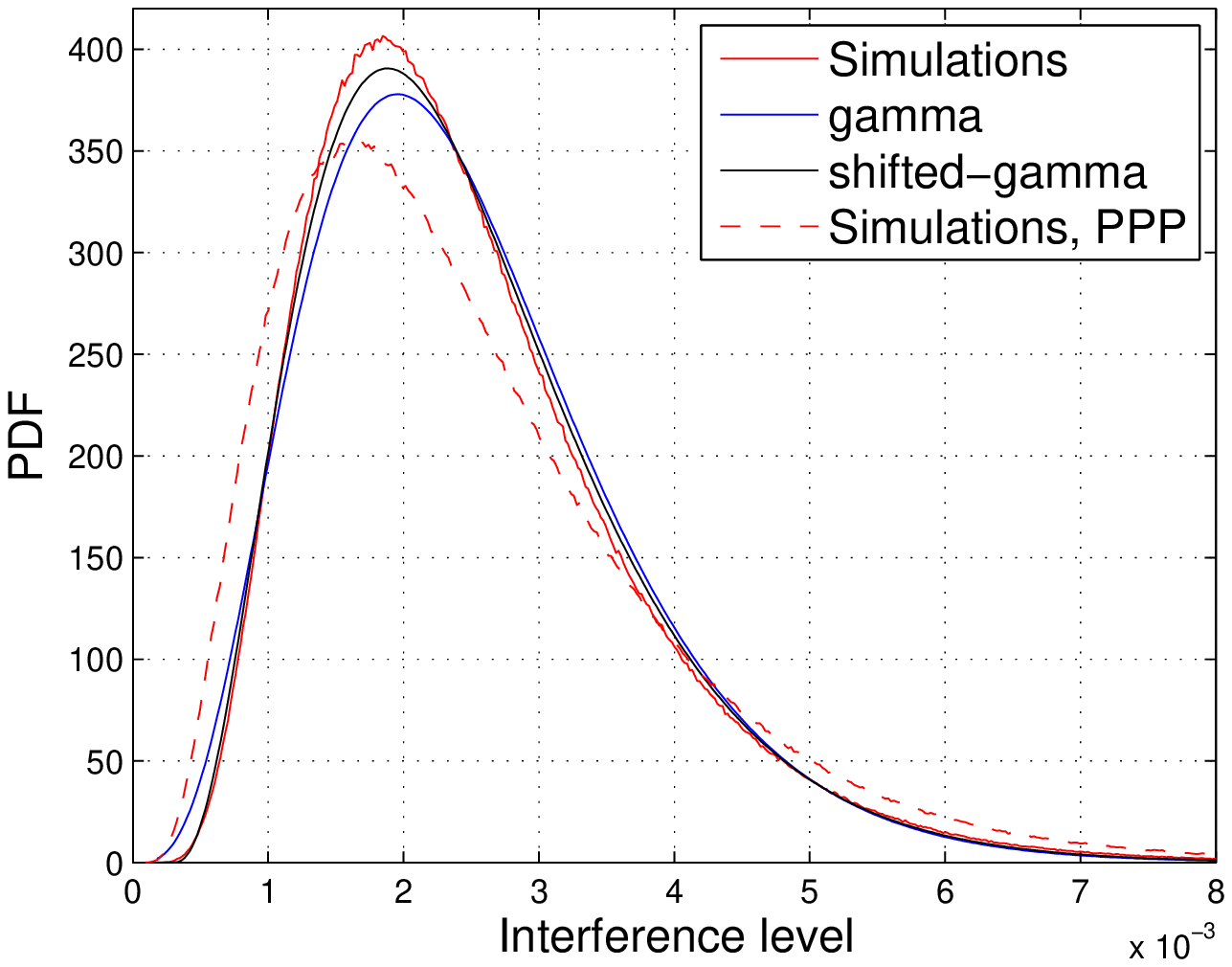}}
 \caption{Simulated \ac{PDF} of the interference level for a hardcore point process along with gamma and shifted-gamma approximations. $r_0\!=\! 100$ m, $\eta\!=\! 3$ and $\lambda c\!=\!0.4$. For the high-speed lane of a motorway in the busy hour $\lambda c\!\in\!\left(0.35,0.45\right)$~\cite[Fig.~8]{Koufos2019}. For fixed $\lambda c$, a higher intensity of vehicles paired with a lower tracking distance can be associated with driving at lower speeds. A \ac{PPP} with equal intensity is also simulated. $10^7$ trials per simulation curve.}
 \label{fig:InterfDistr}
\end{figure*}

The approximation accuracy of the gamma and the shifted-gamma \acp{PDF} is illustrated in Fig.~\ref{fig:InterfDistr} for two intensities $\lambda$, and $\lambda c\!=\!0.4$. The simulated standard deviation and skewness, along with their approximations, are included in Table~\ref{table1}. We see in the table that: (i) the variance approximation in~\eqref{eq:VarApp} is quite accurate, (ii)  Lemma~\ref{lemma:1} estimates the skewness better than the gamma distribution, $\frac{2}{\sqrt{k}}$, and (iii) the \ac{PPP} has higher variance and skewness than the hardcore process, justifying the approximations  in~\eqref{eq:VarApp} and Lemma~\ref{lemma:1}. In both cases, the \ac{PPP} estimates the skewness (in an absolute sense) better than Lemma~\ref{lemma:1} because the values of $\lambda c$ and $\frac{c}{r_0}$ are not close to zero where Lemma~\ref{lemma:1} holds. Nevertheless, the \ac{PPP} gives much worse estimates for the standard deviation (see Table~\ref{table1}) and the interference distribution (see Fig.~\ref{fig:InterfDistr}) than the gamma approximation.  
\begin{table}[!t]
\caption{Standard deviation and skewness of interference for a hardcore process with $\lambda c\!=\!0.4$ obtained by simulations, and estimated using~\eqref{eq:VarApp} and Lemma~\ref{lemma:1}.}
\label{table1}
\centering
\begin{tabular}{ |l|c|c|c|c| } 
 \hline
  {} & sim. & gamma & shifted-gamma & PPP \\ \hline 
 st. dev, $\lambda=0.1\, {\text{m}}^{-1}$ & $0.0024$ & $0.0023$ & $0.0023$ & $0.0028$\\  \hline 
 skewn., $\lambda=0.1\, {\text{m}}^{-1}$ & $0.60$ & $0.46$ & $0.53$ & $0.66$\\ \hline 
 st. dev, $\lambda=0.025\, {\text{m}}^{-1}$ & $0.0012$ & $0.0012$ & $0.0012$ & $0.0014$\\  \hline 
 skewn., $\lambda=0.025\, {\text{m}}^{-1}$ & $1.27$ & $0.93$ & $1.06$ & $1.32$\\ \hline 
\end{tabular}
\end{table}
For fixed $\lambda c$, the skewness and the \ac{CoV} are both proportional to $\frac{1}{\sqrt{\lambda}}$. We see in Fig.~\ref{fig:InterfDistr} that for higher intensity of vehicles, $\lambda\!=\!0.1$, the interference distribution becomes more concentrated and less skewed, and the gamma approximation provides a very good fit. For lower intensity of vehicles,  $\lambda\!=\!0.025$, the skewness and the \ac{CoV} of interference increase, and three moments clearly provide a better fit than two. Also, note that the poor fit of the \ac{PPP} model near origin will translate to a poor fit in the upper tail of the \ac{SIR} \ac{CDF}, which is associated with the outage probability of high rate transmissions.

\section{Probability of outage} 
\label{sec:Outage}
Under Rayleigh fading the probability of outage at operation threshold $\theta$, ${\text{P}_{\text{out}}}\!\left(\theta\right) \!=\! \mathbb{P}\!\left({\text{SIR}}\!\leq\!\theta\right)$, can be written in terms of the \ac{LT} of the interference distribution. Even though the interference \ac{PDF} is unknown, its \ac{LT} could be computed provided that the \ac{PGFL} of the hardcore process was available. Unfortunately, this is not the case. In addition, the bounds of~\cite[Theorem 2.1]{Stucki2014} using the first-order expansion of the \ac{PGFL} and the conditional Papangelou intensity$\!\!\!$~\footnote{Without presenting the calculation details, the upper bound to the \ac{PGFL} (lower bound to the outage probability) follows by setting the local stability constraint in~\cite[equation (2.8)]{Stucki2014} equal to $c^*\!=\!\mu e^{\mu c}$.} are tight only in the lower tail of the outage probability. In order to appoximate the outage probability, we may use the \ac{PPP} of intensity $\lambda$ as an approximation to the hardcore process. 
\begin{equation}
\label{eq:LowerBound}
\begin{array}{ccl}
{\text{P}_{\text{out}}^{\text{ppp}}}\!\left(\theta\right) \!\!\!&\stackrel{(a)}{=}&\!\!\! \displaystyle  1 \!-\!  \mathbb{E}_{\Phi}\left\{\prod\nolimits_{x_k\in\Phi}\frac{1}{1\!\!+\!s\, x_k^{-\eta}} \right\} \\ \!\!\!&\stackrel{(b)}{=}&\!\!\! \displaystyle 1\!-\!e^{-2 \lambda\int_{r_0}^\infty\left(1-\frac{1}{1+s x^{-\eta}}\right){\rm d}x}, 
\end{array}
\end{equation}
where $s\!=\!\frac{\theta }{P_r}$, $(a)$ follows from exponentially \ac{i.i.d.} fading channels, $(b)$ from the \ac{PGFL} of \ac{PPP}, and the integral in the exponent can be expressed in terms of the ${}_2F_1$ Gaussian hypergeometric function~\cite[p.~556]{Abramo}.

An upper bound to the outage probability can also be obtained as follows: 
\[ 
\begin{array}{ccl}
{\text{P}_{\text{out}}}\!\left(\theta\right) \!\!\!&=&\!\!\! \displaystyle 1 - 
\mathbb{E}_{\Phi}\left\{ e^{-\sum\nolimits_{x_k\in\Phi}\log\left(1+s x_k^{-\eta}\right)}\right\} \\ \!\!\!&\stackrel{(a)}{\leq}&\!\!\! \displaystyle 1 - \exp\!\left(-\mathbb{E}_{\Phi}\left\{\sum\nolimits_{x_k\in\Phi}\log\left(1\!+\!s x_k^{-\eta}\right)\right\}\right) \\ \!\!\!&\stackrel{(b)}{=}&\!\!\! \displaystyle 1-\exp\!\left(-2\lambda\!\!\int_{r_0}^\infty\!\!\!\log\!\left(1\!+\!s x^{-\eta}\right){\rm d}x\right) \!=\! {\text{P}_{\text{out}}^{\text{Jen}}}\!\left(\theta\right),  
\end{array}
\]
where $(a)$ is due to Jensen's inequality, $(b)$ follows from Campbell's theorem and the integral in the exponent can be expressed in terms of the ${}_2F_1$ function. 
\begin{figure*}[!t]
 \centering \subfloat[microcell, $r_0=50$ m, $\lambda c\!=\!0.5$] {\includegraphics[width=\columnwidth]{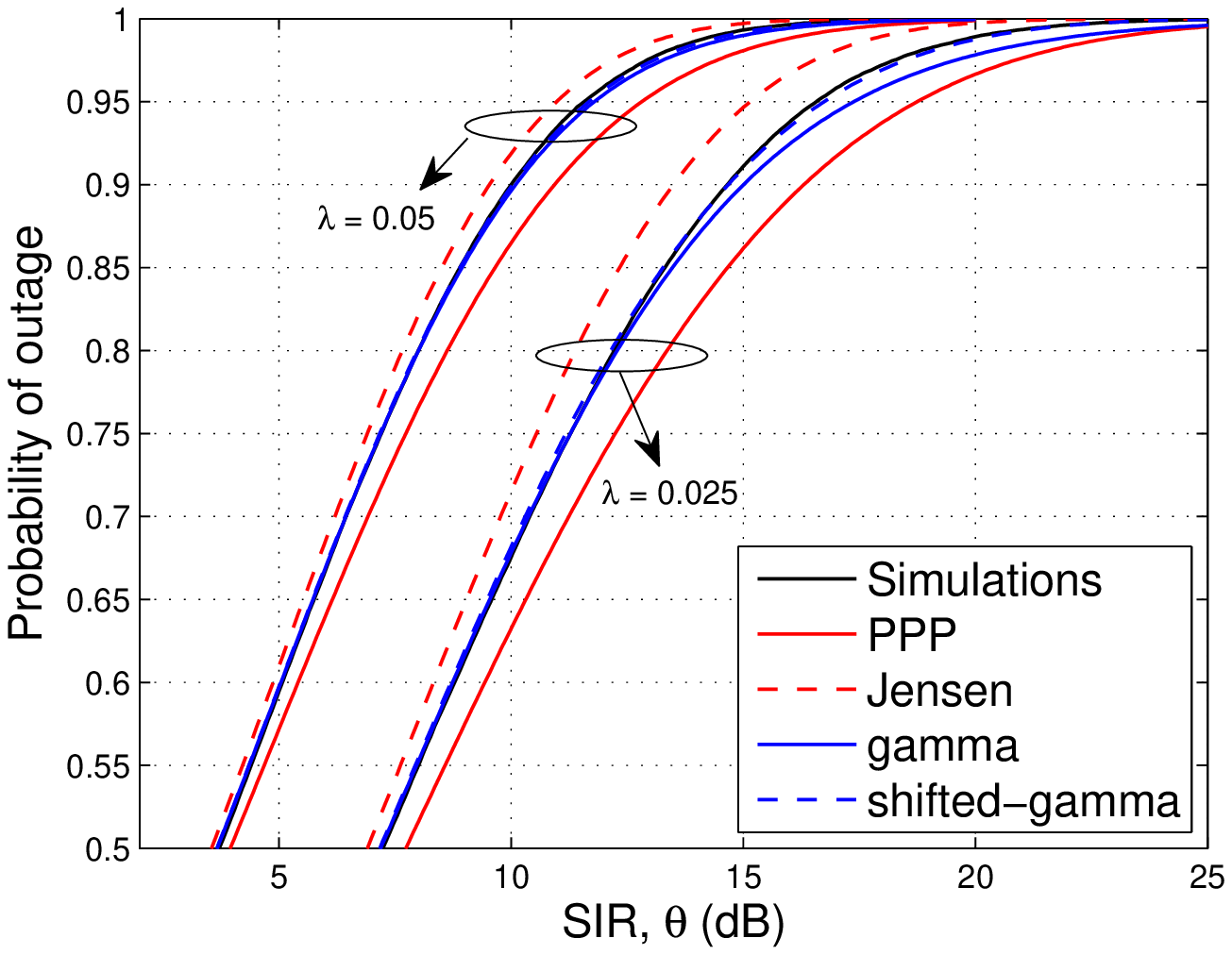}\label{fig:JensenPPPGammasA}}\hfil \subfloat[macrocell, $r_0=250$ m, $\lambda c\!=\!0.4$] {\includegraphics[width=\columnwidth]{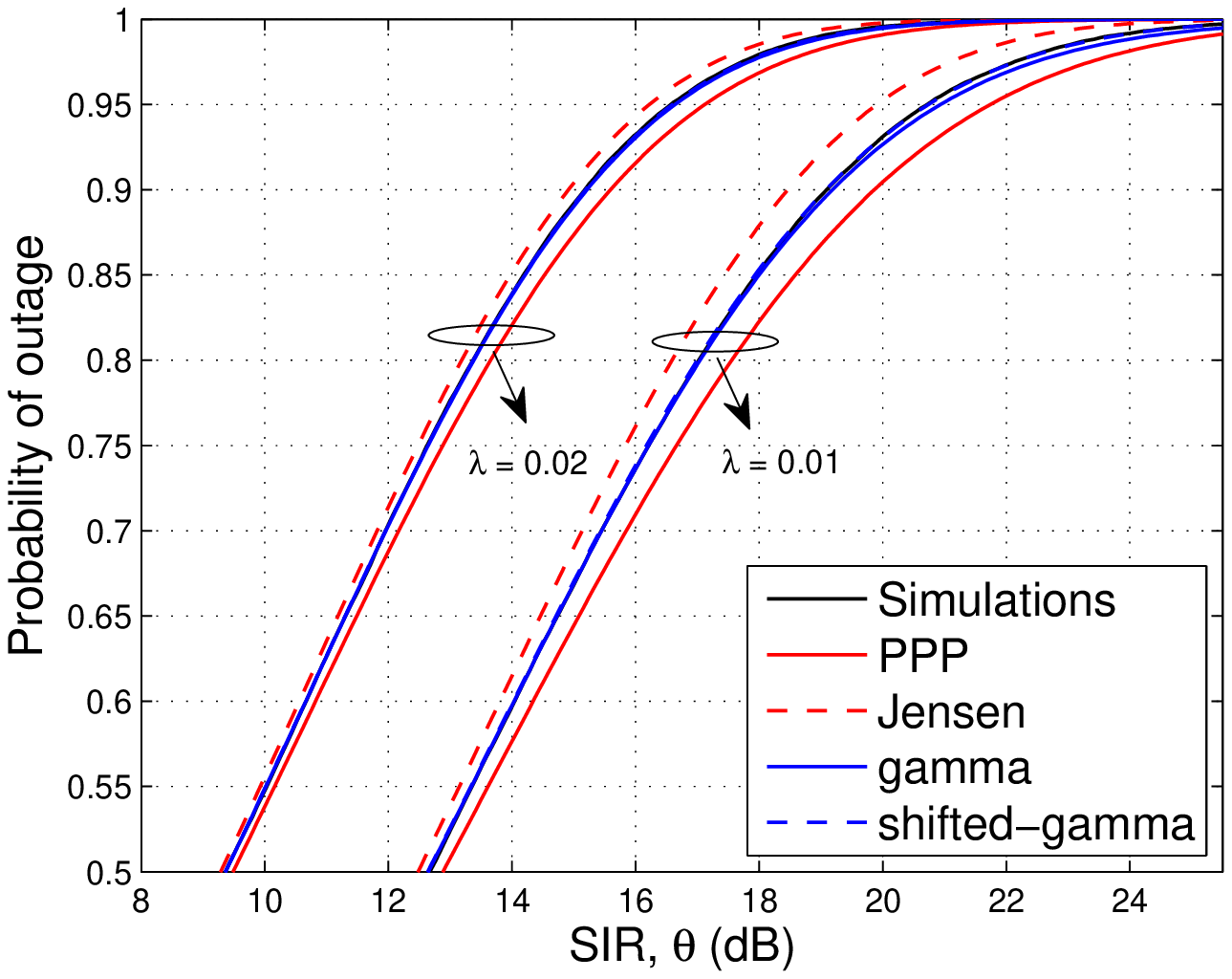}}
 \caption{Simulated probability of outage for the hardcore point process along with the upper-bound, ${\text{P}_{\text{out}}^{\text{Jen}}}\!\left(\theta\right)$, the \ac{PPP} approximation ${\text{P}_{\text{out}}^{\text{ppp}}}\!\left(\theta\right)$, and the approximations, ${\text{P}_{\text{out}}^{\text{g}}}\!\left(\theta\right)$ and ${\text{P}_{\text{out}}^{\text{sg}}}\!\left(\theta\right)$ in~\eqref{eq:OutageGamma}. $10^5$ simulations per curve. Pathloss exponents $\eta\!=\!4$, signal level $P_r\!=\!8\times 10^{-7}{\text{W}}$ for $r_0\!=\! 50$ m and $P_r\!=\!10^{-8}{\text{W}}$ for $r_0\!=\!250$ m. The gamma approximations fit very well the simulations in the lower tail too, see also Fig~\ref{fig:MRC}.}
 \label{fig:JensenPPPGammas}
\end{figure*}

Finally, the gamma approximations for the \ac{PDF} of interference studied in the previous section have simple \acp{LT}, and they can be used to generate simple approximations for the outage probability. 
\begin{equation}
\label{eq:OutageGamma}
\begin{array}{ccl}
{\text{P}_{\text{out}}}\!\left(\theta\right) \!\!\!&\approx&\!\!\! \displaystyle 1 - \left(1+s \beta\right)^{-k} = {\text{P}_{\text{out}}^{\text{g}}}\!\left(\theta\right),  \\ 
{\text{P}_{\text{out}}}\!\left(\theta\right) \!\!\!&\approx&\!\!\! \displaystyle 1-e^{-s \epsilon} \left(1+s\beta\right)^{-k} = {\text{P}_{\text{out}}^{\text{sg}}}\!\left(\theta\right).
\end{array}
\end{equation}

We see in Fig.~\ref{fig:JensenPPPGammas} that the \ac{PPP} and the Jensen inequality are tight in the body of the \ac{SIR} \ac{CDF}, but they start to fail in the upper tail. The error is more prominent in microcells and macrocells with a low intensity of vehicles. Recall from Section~\ref{sec:Distribution} that smaller cell sizes $r_0$ and lower intensities $\lambda$ are associated with higher \ac{CoV} and skewness for the interference distribution. According to~\eqref{eq:VarApp} and Lemma~\ref{lemma:1}, for a fixed $\lambda c$, the absolute prediction error of \ac{PPP} increases for lower $\lambda,r_0$, and subsequently, the induced errors for the interference distribution and the outage probability become higher. We claim that the \ac{PPP} cannot always describe accurately the outage probability of a link in a field of interferers with hardcore headway distance. We will illustrate next that for temporal performance metrics and multiple antennas at the receiver the \ac{PPP} accuracy worsens, while the gamma approximations can be used to generate good performance predictions in all cases. 

\section{Applications}
\label{sec:Applications}
The two deployment models (hardcore vs. \ac{PPP}) induce different interference correlation over time and space. We will use the mean local delay to describe the temporal performance of the link, and a dual-branch \ac{MRC} receiver for the spatial performance. For notational brevity, we will use the gamma approximation for the distribution of interference, unless otherwise stated.

\subsection{Temporal performance}
The mean local delay is defined as the average number of transmissions required for successful reception. For mobility models introducing correlations in the locations of interferers over time, it is challenging to calculate it. For $T$ consecutive transmissions, the joint $T-$th dimensional \ac{PDF} of interference, with correlated marginals, would be needed. Alternatively, we may get some insight by investigating the properties of delay under (i) i.i.d. locations, and (ii) static interferers over time~\cite{Haenggi2013b}. The performance is associated with scenarios characterized by very high and very low mobility of interferers respectively. 

For \ac{i.i.d.} locations, the mean delay is equal to the inverse of the probability of successful reception. For the \ac{PPP}, one may take the complementary of the last line of~\eqref{eq:LowerBound} and invert it. For the hardcore process, the mean delay would be approximated by $\left(1\!+\!s \beta\right)^k$, see~\eqref{eq:OutageGamma}. 

In order to calculate the mean delay with static interferers, one has to invert the probability of successful reception conditioned on the realization of interferers, then average over their locations~\cite{Haenggi2013b}. The mean delay with Poisson interferers accepts an elegant form for continuous transmissions, $\mathbb{E}\!\left\{D\right\}\!=\!e^{s\,  \mathbb{E}\left\{\mathcal{I}\right\}}$, which follows from substituting $p\!=\!1, q\!=\!0$ in~\cite[Lemma 2]{Haenggi2013b}. In order to overcome the lack of the \ac{PGFL} for the hardcore process, we use an alternative expression for the mean delay, $\mathbb{E}\!\left\{D\right\}\!=\!\sum\nolimits_{T=1}^\infty \!\! {\text{P}_{\text{out}}}\!\left(T\right)$~\cite[Section V-B]{Haenggi2013c}, where ${\text{P}_{\text{out}}}\!\left(T\right)$ is the joint outage probability over $T$ consecutive time slots. 
\[
\begin{array}{ccl}
\text{P}_{\text{out}}\!\left(T\right) \!\!\!\!\!&=&\!\!\!\!\! \displaystyle  \mathbb{P}\!\left(\text{SIR}_1\!\leq\!\theta,\, \text{SIR}_2\!\leq\!\theta,\ldots, \, \text{SIR}_T\!\leq\!\theta\right) \\ \!\!\!\!\!&=&\!\!\!\!\! \displaystyle \mathbb{E}\left\{\left(1\!-\!e^{-s \mathcal{I}_1}\right)\left(1\!-\!e^{-s \mathcal{I}_2}\right)\ldots \left(1\!-\!e^{-s \mathcal{I}_T}\right)\right\} \\ \!\!\!\!\!&=&\!\!\!\!\!  \displaystyle \sum\limits_{t=0}^T \left(-1\right)^t C^T_t  \mathbb{E}\!\left\{ e^{-s \sum\nolimits_{j=1}^t \mathcal{I}_j}\right\} \\ \!\!\!\!\!&=&\!\!\!\!\! \displaystyle \sum\limits_{t=0}^T \left(-1\right)^t C^T_t \mathbb{E}\!\left\{ e^{-s \sum\nolimits_{x_k\in\Phi}\sum\nolimits_{j=1}^t h_{k,j}g\left(x_k\right)}\right\} \\ \!\!\!\!\!&=&\!\!\!\!\!  \displaystyle  \sum\limits_{t=0}^T \left(-1\right)^t C^T_t \mathbb{E}\!\left\{ e^{-s \sum\nolimits_{x_k\in\Phi} h_k\left(t\right)g\left(x_k\right)}\right\},
\end{array}
\]
where $C^T_t$ are the $t$-combinations in a $T$-element set,  $\text{SIR}_j$ and $\mathcal{I}_j$ describe the \ac{SIR} and the interference respectively over the $j$-th time slot, and the \ac{RV} $h_k\!\left(t\right)\!=\!\sum_{j=1}^t h_{k,j}$, as a sum of \ac{i.i.d.} exponential \acp{RV} follows the gamma distribution. 

We deduce that the calculation of the joint Laplace functional over $t$ slots with static interferers is equivalent to the calculation of the \ac{LT} of interference for a single time instance, but with a different fading distribution. The first two moments of the \ac{RV} $h_k\!\left(t\right)\!=\!h\!\left(t\right)\, \forall k$ are  $\mathbb{E}\!\left\{h\!\left(t\right)\right\} \!=\! t$ and   $\mathbb{E}\!\left\{h^2\!\left(t\right)\right\} \!=\! t\left(1\!+\!t\right)$. We will still utilize the gamma approximation for the interference, but the fading is now modeled by a gamma instead of an exponential \ac{RV}. We have the same simple expression for the \ac{LT} of interference, $\left(1\!+\!s\beta\!\left(t\right)\right)^{-k\left(t\right)}$, where the parameters $k,\beta$ now depend on $t$. Without showing the derivation details, the mean, and the  variance of interference in the presence of Nakagami fading modeled by a gamma \ac{RV} with shape $t$ and scale unity are 
\begin{equation}
\label{eq:MeanVarT}
\begin{array}{ccl}
\mathbb{E}\!\left\{ \mathcal{I}\!\left(t\right) \right\} \!\!\!\!\!&=&\!\!\!\!\! \displaystyle \frac{2\lambda \, t \, r_0^{1-\eta}}{\eta-1} \\
\mathbb{V}\!\left\{ \mathcal{I}\!\left(t\right) \right\} \!\!\!\!\!&\approx&\!\!\!\!\! \displaystyle \frac{2\lambda t \,  (1\!+\!t\left(1\!-\!\lambda c\right)^2) \, r_0^{1-2\eta}}{2\eta\!-\!1}.
\end{array}
\end{equation}

Finally, the approximation for the mean delay using the gamma distribution can be read as  
\[
\begin{array}{ccl}
\mathbb{E}\!\left\{D\right\} \!\!\!\!\!\!&\approx&\!\!\!\!\!\! \displaystyle \sum_{T=0}^\infty  \sum_{t=0}^T \left(-1\right)^t C^T_t \left(1+s \beta\!\left(t\right)  \right)^{-k\left(t\right)}, 
\end{array}
\]
where $k\!\left(t\right), \beta\!\left(t\right)$ are derived via moment matching using~\eqref{eq:MeanVarT}. 

The above approximation can be turned into a single summation by re-ordering the two sums and setting a sufficient maximum value $T_0$ for the parameter $T$, where the sum over $t$ should be truncated. 
\begin{equation}
\label{eq:MeanDelayGammas2}
\begin{array}{ccl}
\mathbb{E}\!\left\{D\right\} \!\!\!\!\!&=&\!\!\!\!\! \displaystyle \sum_{t=0}^\infty  \sum_{T=t}^\infty \left(-1\right)^t C^T_t \left(1+s \beta\!\left(t\right)  \right)^{-k\left(t\right)} \\ \!\!\!\!\!&=&\!\!\!\!\! \displaystyle  \lim_{T_0\!\rightarrow\!\infty} \sum_{t=0}^{T_0}  \sum_{T=t}^{T_0} \left(-1\right)^t C^T_t \left(1+s \beta\!\left(t\right)  \right)^{-k\left(t\right)}  \\ \!\!\!\!\!&=&\!\!\!\!\! \displaystyle \lim_{T_0\!\rightarrow\!\infty} \sum\limits_{t=0}^{T_0}  \left(-1\right)^t C^{T_0+1}_{t+1} \left(1+s \beta\!\left(t\right)  \right)^{-k\left(t\right)}.
\end{array}
\end{equation}
\begin{figure}[!t]
 \centering
\includegraphics[width=\linewidth]{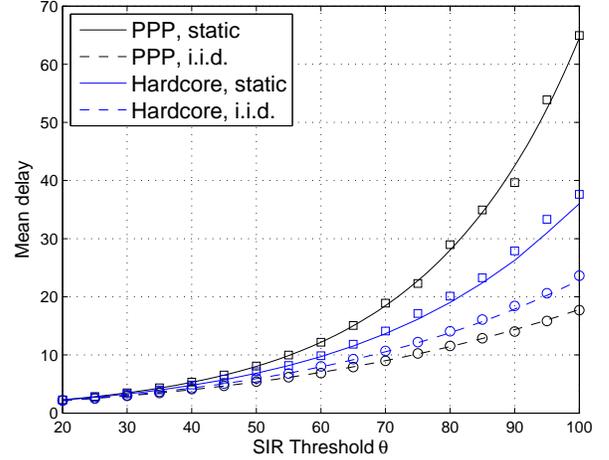} 
 \caption{Mean local delay at the origin under different fields of interferers. $r_0\!=\!100 \, \text{m}, \eta\!=\!4, \lambda\!=\!0.05\, {\text{m}}^{-1}, c\!=\!8\, {\text{m}}, P_r\!=\!8\times 10^{-6} \text{W}$. In the numerical evaluation of~\eqref{eq:MeanDelayGammas2}, we truncated at  $T_0\!=\!5\,000$ and used $2\, 000$-digit precision in Mathematica~\cite{Mathematica}. We validated numerically that higher values of $T_0$ give negligible additional contribution to the limit. $10\, 000$ simulations per marker. Solid and dashed lines use the gamma approximation for the hardcore process and they are exact calculations for the \ac{PPP}.}
 \label{fig:MeanDelay}
\end{figure}

Since it is not realistic to assume very low mobility across macrocells, we depict in Fig.~\ref{fig:MeanDelay} the mean delay for a microcell. The interference field due to the hardcore process induces a much smaller increase in the mean delay in comparison with \ac{PPP}, as we move from extreme mobile to static interferers. This is because the temporal correlation coefficient of interference due to a static \ac{PPP} is equal to $\frac{1}{2}$~\cite{Haenggi2009}, while that due to the hardcore process is approximately $\frac{1}{2}\left(1\!-\!\lambda c\right)$~\cite{Koufos2018b}. Due to the lower correlation of interference, fewer retransmissions are needed (on average) to meet the \ac{SIR} target, and the mean delay decreases in comparison with that due to static $\ac{PPP}$. 

\subsection{Spatial performance}
The probability of successful reception for \ac{MRC} with dual-branch receiver has been derived in~\cite[equation (26)]{Tanbourgi2014} for a Poisson field of interferers. The \ac{PGFL} for the hardcore process is not known and thus, we resort  again to approximations about the distribution of interference in the two branches and their correlation. We will end up with a simple approximation for the outage probability, while the calculation in~\cite[equation (26)]{Tanbourgi2014} requires the numerical computation of three integrals.

Let us denote by $\mathcal{I}_1\!=\!\sum\nolimits_i h_{1,i}g\left(x_i\right)$ and $\mathcal{I}_2\!=\!\sum\nolimits_i h_{2,i}g\left(x_i\right)$, the instantaneous interference, and by $\mathbf{I}$ the vector of $\mathcal{I}_1,\mathcal{I}_2$. Treating the interference as white noise, the \ac{MRC} becomes optimal, and the post-combining \ac{SIR} equals the sum of the \acp{SIR} at the two branches. 
\[
\mathbb{P}\!\left\{\text{SIR}\geq\theta\right\} = \mathbb{E}_{\mathbf{I}}\!\left\{ \mathbb{P}\left( \frac{h_{t,1}P_r}{\mathcal{I}_1}+\frac{h_{t,2}P_r}{\mathcal{I}_2}\geq\theta|\mathbf{I}\right)\right\}. 
\]

Let us denote by $W\!=\!\frac{h_{t,2}P_r}{\mathcal{I}_2}$ the \ac{RV} describing the \ac{SIR} at the second branch. Conditioning on the realization $w$, and using that the fading channel is Rayleigh, we have 
\[
\begin{array}{ccl}
\mathbb{P}\!\left\{\text{SIR}\geq\theta\right\} \!\!\!\!\!\!&=&\!\!\!\!\! \displaystyle \mathbb{E}_{\mathbf{I},W}\!\left\{e^{-s_1\mathcal{I}_1}\right\} \!=  \mathbb{E}_{\mathbf{I}}\!\left\{\! \int_0^\infty \!\!\!\!\!\!  e^{-s_1\mathcal{I}_1} f_{W|\mathcal{I}_2}\!\left(w\right) \!{\rm d}w \right\}\!, 
\end{array}
\]
where $s_1\!=\!\frac{\max\left\{0,\theta-w\right\}}{P_r}$ and $f_{W|\mathcal{I}_2}$ is the conditional \ac{PDF} of the \ac{SIR} at the second branch. 

Due to the fact that the fading channel is Rayleigh, $\mathbb{P}\left(W\!\geq\!w|\mathcal{I}_2\right)=e^{-s_2 \mathcal{I}_2}$, where $s_2\!=\!\frac{w}{P_r}$. By differentiation,  $f_{W|\mathcal{I}_2}\!\left(w\right)\!=\!\frac{\mathcal{I}_2}{P_r}e^{-s_2 \mathcal{I}_2}$. Therefore 
\begin{equation}
\label{eq:MRC}
\begin{array}{ccl}
\mathbb{P}\!\left\{\text{SIR}\geq\theta\right\} \!\!\!\!\!\!&=&\!\!\!\!\!\! \displaystyle \frac{1}{P_r} \int_0^\infty  \mathbb{E}_{\mathbf{I}}\!\left\{ \mathcal{I}_2 e^{-s_1\mathcal{I}_1} e^{-s_2 \mathcal{I}_2} \right\} {\rm d}w \\ \!\!\!\!\!\!&\stackrel{(a)}{=}&\!\!\!\!\!\! \displaystyle \frac{1}{P_r} \int_0^\theta  \mathbb{E}_{\mathbf{I}}\!\left\{ \mathcal{I}_2 e^{-s_1\mathcal{I}_1} e^{-s_2 \mathcal{I}_2} \right\} {\rm d}w \, + \\ & & \displaystyle \frac{1}{P_r} \int_\theta^\infty \mathbb{E}_{\mathbf{I}}\!\left\{ \mathcal{I}_2 e^{-s_2 \mathcal{I}_2} \right\} {\rm d}w, 
\end{array}
\end{equation}
where $(a)$ follows from $s_1\!=\!0$ for $w\!>\!\theta$.

We will assume that the random vector $\mathbf{I}$ follows the bivariate gamma distribution with identical marginals following the gamma distribution with parameters $\left\{k,\beta\right\}$ calculated in Section~\ref{sec:Distribution}. The correlation coefficient is denoted by $\rho$, and it is calculated in the Lemma~\ref{lemma:2} below. Using the differentiation property of the \ac{LT}~\cite[pp.~229]{Kobayashi2012}, the first expectation in~\eqref{eq:MRC}, $\mathcal{J}\!=\!\mathbb{E}_{\mathbf{I}}\!\left\{ \mathcal{I}_2 e^{-s_1\mathcal{I}_1} e^{-s_2 \mathcal{I}_2} \right\}$, becomes  
\begin{equation}
\label{eq:LapApp1}
\begin{array}{ccl}
\mathcal{J} \!\!\!&=&\!\!\! \displaystyle -\frac{\partial}{\partial s_2}\left\{\left(1\!+\!s_1\beta\!+\!s_2\beta\!+\!s_1s_2\beta^2\left(1\!-\!\rho\right) \right)^{-k}\right\} \\ \!\!\!&=&\!\!\! \displaystyle \frac{k\beta\left(1+\beta s_1\left(1-\rho\right)\right)}{ \left(1+s_1\beta+s_2\beta+s_1s_2\beta^2\left(1-\rho\right) \right)^{k+1}}.
\end{array}
\end{equation}

\noindent 
The second expectation in~\eqref{eq:MRC} is 
\begin{equation}
\label{eq:LapApp2}
\mathbb{E}_{\mathbf{I}}\!\left\{ \mathcal{I}_2 e^{-s_2 \mathcal{I}_2} \right\} = k\beta\left(1+s_2\beta\right)^{-k-1}.
\end{equation}

After substituting~\eqref{eq:LapApp1} and~\eqref{eq:LapApp2} into~\eqref{eq:MRC}, cancelling out some terms and carrying out the integration for $w\!>\!\theta$, we end up with 
\begin{equation}
\label{eq:MRC2}
\begin{array}{ccl}
\mathbb{P}\!\left\{\text{SIR}\geq\theta\right\} \!\!\!&=&\!\!\! \displaystyle P_r^k \left(P_r+\theta\beta\right)^{-k} + k\beta P_r^{2k}\, \times  \\ \!\!\!\!\!\!\!\!\!\!\!\!\!\!\!\!\!\!\!\!\!\!\!& &\!\!\!\!\!\!\!\!\!\!\!\!\!\!\!\!\!\!\!\!\!\! \displaystyle \int_0^\theta \!\!\!\! \frac{\left(P_r+\beta \left(\theta-w\right)\left(1-\rho\right)\right) {\rm d}w}{\left(P_r^2 \!+\! \theta\beta P_r \!+\! \left(\theta-w\right)w\beta^2\left(1\!-\!\rho\right) \right)^{k+1}}. \\ 
\end{array}
\end{equation}

\noindent 
The above integral can be expressed in terms of ${}_2F_1$ functions. 

\begin{figure}[!t]
 \centering 
\includegraphics[width=\linewidth]{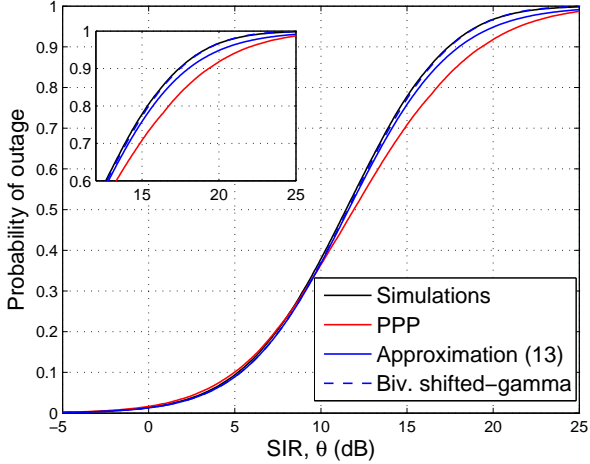}
 \caption{Probability of outage at the origin with dual-antenna \ac{MRC} and interferers deployed as a hardcore process. The approximation~\eqref{eq:MRC2} is verified with simulations. $\lambda\!=\!0.025 {\text{m}}^{-1}, c\!=\! 20 \text{m}$. See the caption of Fig.~\ref{fig:JensenPPPGammas} for the rest of the parameter settings for microcells. In the inset we zoom at the upper tail. The interference approximation using a bivariate shifted-gamma distribution with correlated marginals gives an excellent fit to the simulations.}
 \label{fig:MRC}
\end{figure}

In Fig.~\ref{fig:MRC} we depict the outage probability. The performance prediction of \ac{PPP} in the upper tail worsens in comparison with single-antenna receiver, and it is expected to deteriorate with more antennas. Without presenting the calculation details, the approximation for the outage probability using a bivariate shifted-gamma distribution for the distribution of interference at the two branches of the receiver is also included in Fig.~\ref{fig:MRC}. The two approximations can be used to get a quite good performance estimate with low computational complexity. 
\begin{lemma}
\label{lemma:2}
For \ac{i.i.d.} exponential power fading channels with unit mean, the spatial correlation coefficient of interference $\rho$ between the two antennas can be approximated as  $\rho\!\approx\!\frac{1}{2}\left(1-\lambda c\right)$. The approximation is valid for $\lambda c\!\rightarrow\!0$ and $\frac{c}{r_0}\!\rightarrow\!0$. 
\begin{proof}
The covariance of interference is 
\[
\begin{array}{ccl}
{\text{cov}}\!\left\{\mathcal{I}\right\} \!\!\!\!\!&=&\!\!\!\!\! \displaystyle  \mathbb{E}\!\left\{h\right\}^{\!2}  \mathbb{E}\!\left\{\!\sum\nolimits_{x\in\Phi}\! g^2\!\left(x\right)\!\right\} + \\ & & \displaystyle \mathbb{E}\!\left\{h\right\}^{\!2} \mathbb{E}\!\left\{\!\sum\nolimits_{x,y\in\Phi}^{x\neq y} \! g\!\left(x\right) g\!\left(y\right)\!\right\} - \mathbb{E}\!\left\{\mathcal{I}\right\}^{\!2} \\  \!\!\!\!\!&=&\!\!\!\!\! \displaystyle   \frac{2\lambda r_0^{1-2\eta}}{2\eta-1} \!+\!\! \int \!\! g\!\left(x\right) g\!\left(y\right) \rho^{\left(2\right)}\!\left(x,y\right) {\rm d}x {\rm d}y \!-\!  \mathbb{E}\!\left\{\mathcal{I}\right\}^{\!2}\!\!. 
\end{array}
\]

\noindent
The variance of interference is~\cite[equation (3)]{Koufos2018}
\[
\begin{array}{ccl}
{\text{V}}\!\left\{\mathcal{I}\right\} \!\!\!\!\!&=&\!\!\!\!\! \displaystyle  \mathbb{E}\!\left\{h^2\right\} \mathbb{E}\!\left\{\sum\nolimits_{x\in\Phi} g^2\!\!\left(x\right)\!\right\} + \\ & & \displaystyle  \mathbb{E}\!\left\{h\right\}^{\!2}\mathbb{E}\!\left\{\!\sum\nolimits_{x,y\in\Phi}^{x\neq y} \! g\!\left(x\right) g\!\left(y\right)\!\!\right\} - \mathbb{E}\!\left\{\mathcal{I}\right\}^{\!2} \\  \!\!\!\!\!&=&\!\!\!\!\! \displaystyle   \frac{4\lambda r_0^{1-2\eta}}{2\eta-1} + \int \! g\!\left(x\right) g\!\left(y\right) \rho^{\left(2\right)}\!\left(x,y\right) {\rm d}x {\rm d}y \!-\!  \mathbb{E}\!\left\{\mathcal{I}\right\}^{\!2}\!. 
\end{array}
\]

The integral $S\!=\!\int g\!\left(x\right) g\!\left(y\right) \rho^{\left(2\right)}\!\left(x,y\right) {\rm d}x {\rm d}y$ has been approximated in~\cite[Section V]{Koufos2018} for $\lambda c\!\rightarrow\! 0$ and $\frac{c}{r_0}\!\rightarrow\! 0$. The two dominant terms with respect to $r_0$ are 
\[
S \approx \frac{4\lambda^2 r_0^{2-2\eta}}{\left(\eta-1\right)^2} - \frac{4\lambda^2 c r_0^{1-2\eta}}{2\eta-1} + \frac{2\lambda^3 c^2 r_0^{1-2\eta}}{2\eta-1}. 
\]

After substituting the above approximation for $S$ in the expressions of the covariance and the variance, doing some factorization and cancelling out common terms, the correlation coefficient can be approximated as 
\[
\rho = \frac{{\text{cov}}\!\left\{\mathcal{I}\right\}}{{\text{V}}\!\left\{\mathcal{I}\right\}} \approx \frac{\left(1-\lambda c\right)^2}{2-2\lambda c + \lambda^2 c^2} \stackrel{\lambda c \rightarrow 0}{\approx} \frac{1}{2}\left(1-\lambda c\right),
\]
and the Lemma is proved.
\end{proof}
\end{lemma}

\section{Conclusions}
\label{sec:Conclusions}
The \ac{PPP} model for vehicular networks allows small inter-vehicle distances with non-negligible probability. A more realistic point process of equal intensity, but with a hardcore distance, i.e., shifted-exponential inter-arrivals~\cite{Koufos2019}, changes the properties of interference distribution at the origin. The discrepancy in the outage probability predicted by the two models is evident when the coefficient of variation and the skewness of interference is high, i.e., in urban microcells and motorway macrocells with sparse flows. The discrepancy increases if we consider multiple antennas at the receiver because, in that case, the outage probability also depends on the spatial correlation of interference, which is different under the two deployment models. Temporal performance indicators associated with the performance of retransmission schemes are affected by the correlation properties of interference too. We bypassed the lack of the \ac{PGFL} for the hardcore point process by using a gamma approximation for the interference distribution. The gamma distribution yields better predictions for the outage probability than the \ac{PPP} in all cases. In this paper, we assumed that the point process impacts the distribution of interferers while the transmitter-receiver link is fixed and known. It would be interesting to use a random link distance and investigate whether the horizontal deployment gain for non-Poisson stationary point processes holds~\cite{Haenggi2015}. Another promising topic is the validation of the models against real vehicular traces~\cite{Koufos2019b}. 

\section*{appendix}
\setcounter{equation}{0}
\renewcommand{\theequation}{A.\arabic{equation}}

The third moment of interference accepts contributions from a single user, from user pairs and from triples of users. 
\begin{equation}
\begin{array}{ccl}
\mathbb{E}\!\left\{\mathcal{I}^3\right\} \!\!\!\!\!&=&\!\!\!\!\! \displaystyle  \mathbb{E}\!\left\{\!h^3\!\right\}\!\lambda\!\! \int \!\!\!g^3\!\left(x\right) {\rm d}x \, + \\ & & \displaystyle 3\mathbb{E}\!\left\{\!h^2\!\right\}\!\!\int\!\!\! g^2\!\left(x\right) g\!\left(y\right) \rho^{(2)}\!\left(x,y\right) \! {\rm d}x{\rm d}y  + \\ & & \displaystyle  \int\!\!\! g\!\left(x\right)g\!\left(y\right)g\!\left(z\right) \rho^{(3)}\!\left(x,y,z\right) \! {\rm d}x{\rm d}y{\rm d}z. 
\end{array}
\end{equation}

\noindent 
After substituting the moments of Rayleigh fading we get 
\begin{equation}
\label{eq:ThirdMomInterf}
\begin{array}{ccl}
\mathbb{E}\!\left\{\mathcal{I}^3\right\} \!\!\!\!\!&=&\!\!\!\!\! \displaystyle 6 \lambda\!\! \int \!\!\!g^3\!\left(x\right) \!{\rm d}x  \!+\!6\!\!\int\!\!\! g^2\!\left(x\right)\! g\!\left(y\right) \rho^{(2)}\!\left(x,y\right) \! {\rm d}x{\rm d}y  + \\ & & \displaystyle \int\!\!\! g\!\left(x\right)\! g\!\left(y\right) \! g\!\left(z\right)  \rho^{(3)}\!\left(x,y,z\right) \! {\rm d}x{\rm d}y{\rm d}z \\ \!\!\!&=&\!\!\! \displaystyle \frac{12\lambda r_0^{1-\eta}}{\eta-1}  \!+\!6\!\!\int\!\!\! g^2\!\left(x\right)\! g\!\left(y\right) \rho^{(2)}\!\left(x,y\right) \! {\rm d}x{\rm d}y \, + \\ & & \displaystyle \int\! g\!\left(x\right)\! g\!\left(y\right) \! g\!\left(z\right)  \rho^{(3)}\!\left(x,y,z\right) \! {\rm d}x{\rm d}y{\rm d}z, 
\end{array}
\end{equation}
where $\mathbb{E}\!\left\{h^3\right\}\!=\!6$ and $\mathbb{E}\!\left\{h^2\right\}\!=\!2$ for an exponential \ac{RV}, and we have scaled the second term by three to count the ways to select a user pair out of a triple of users. 

The contributions from triples of users involve the third-order correlation. In order to approximate it, we will apply twice the approximation for the  \ac{PCF} adopted in~\cite{Koufos2018}. Similar to~\cite{Koufos2018}, we will also assume that the guard zone is much larger than the tracking distance, $r_0\!\gg\!c$. In our approximations, we will keep up to the second order terms, and also the dominant $r_0$ terms with exponents larger or equal to $\left(1\!-\!3\eta\right)$.  

Using the approximation for the \ac{PCF} beyond $2c$, the term  $S'\!=\!\iint\!\! x^{-2\eta}y^{-\eta} \rho^{(2)}\!\left(x,y\right) {\rm d}x{\rm d}y$ can be read as 
\begin{equation}
\label{eq:Sprim0}
\begin{array}{ccl}
S' \!\!\!\!\! &\approx& \!\!\!\!\! \displaystyle 2\lambda\mu\int_{r_0}^\infty\!\int_{x+c}^{x+2c}\!\frac{x^{-2\eta}y^{-\eta}}{e^{\mu\left(y-x-c\right)}} {\rm d}y{\rm d}x \, + \\ & & \displaystyle  2\lambda\mu\int_{r_0}^\infty\!\int_{x-2c}^{x-c}\!\frac{x^{-2\eta} g\!\left(y\right)}{e^{\mu\left(x-y-c\right)}} {\rm d}y{\rm d}x \,\, + \\   & &  \displaystyle \,\,\, 2\lambda^2\int_{r_0}^\infty\!\int_{x+2c}^\infty \! x^{-2\eta}y^{-\eta}{\rm d}y{\rm d}x \,\,\,\,\,\, + \\ & & \displaystyle \,\,\,\,\,\, 2\lambda^2\int_{r_0}^\infty\!\int_{-\infty}^{x-2c}\! x^{-2\eta}g\!\left(y\right) {\rm d}y {\rm d}x, 
\end{array}
\end{equation}
where the factor $2$ in front of the integrals accounts for $x\!<\!\!-\!r_0$. 

The contribution to $S'$ due to pairs at distances larger than $2c$ is the last two lines of~\eqref{eq:Sprim0}.
\[
\begin{array}{ccl}
S'_{>2c} \!\!\!\!\!&=&\!\!\!\!\! \displaystyle 2\lambda^2 \!\! \int_{r_0}^\infty\!\!\!\int_{x+2c}^\infty \!\!\! x^{-2\eta}y^{-\eta}{\rm d}y{\rm d}x \,\,\, + \\ \!\!\!\!\!& &\!\!\!\!\! \displaystyle 2\lambda^2 \!\! \int\limits_{r_0}^\infty\!\int\limits_{-\infty}^{-r_0}\!\!\! x^{-2\eta}\left|y\right|^{-\eta}\!{\rm d}y {\rm d}x +\!\!\!\! \int\limits_{r_0+2c}^\infty\!\int\limits_{r_0}^{x-2c}\!\!\! x^{-2\eta}y^{-\eta} {\rm d}y {\rm d}x. 
\end{array}
\]

The first and the last term in the above expression are not equal due to asymmetry in the exponents of $x$ and $y$. After integrating and adding up the three terms we end up with 
\[
\begin{array}{ccl}
S'_{>2c} \!\!\!\!\! &=& \!\!\!\!\! \displaystyle \frac{2\lambda^2\left(r_0^{2-3\eta}\!+\!r_0^{1-\eta}\left(2c\!+\!r_0\right)^{1-2\eta}\right)}{\left(\eta-1\right)\left(2\eta-1\right)} \,\,\, + \\ \!\!\!\!\!\!\!& &\!\!\!\!\!\!\! \displaystyle \frac{2\lambda^2 r_0^{2-3\eta}}{\left(3\eta\!-\!2\right) \left(\eta\!-\!1\right)} \Bigg({}_2F_1\!\left(\!3\eta\!-\!2,\eta\!-\!1,3\eta\!-\!1,\!-\frac{2c}{r_0} \!\right) \!- \\ & & \displaystyle  \,\,\,\,\,\, {}_2F_1\!\left(3\eta\!-\!2,2\eta,3\eta\!-\!1,-\frac{2c}{r_0} \right)\Bigg) \\ \!\!\!\!\!&\stackrel{(a)}{\approx}&\!\!\!\!\!  \displaystyle \frac{4\lambda^2 r_0^{2-3\eta}}{\left(2\eta\!-\!1\right)\left(\eta\!-\!1\right)} - \frac{8 r_0^{1-3\eta} \lambda^2 c}{3\eta-1}, 
\end{array}
\]
where $(a)$ follows from expanding $b\!=\!\frac{c}{r_0}\!\rightarrow\! 0$.

The contribution to $S'$ due to pairs of vehicles at distances $\left|y\!-\!x\right|$ smaller than $2c$ is larger for $y\!>\!x$ (there are no vehicles to filter out inside the cell in that case) than it is for $y\!<\!x$. Nevertheless, for a small $\frac{c}{r_0}$, the two integrals in the first line of~\eqref{eq:Sprim0} should be approximately equal. By making this assumption, we can avoid the approximation of the integral for $y\!<\!x$, which is a little more tedious because for $x\!\in\!\left(r_0,r_0\!+\!2c\right)$ we have to exclude the vehicles inside the cell. Finally, we can approximate the term $S'_{<2c}$ as 
\begin{equation}
\label{eq:Sprim}
\begin{array}{ccl}
S'_{<2c} \!\!\! &\approx& \!\!\! \displaystyle 4\lambda\mu \int_{r_0}^\infty\!\!\int_{x+c}^{x+2c}\!\! \frac{x^{-2\eta}y^{-\eta}}{e^{\mu\left(y-x-c\right)}} {\rm d}y{\rm d}x. 
\end{array}
\end{equation}

\noindent 
After integrating  with respect to $y$  we get 
\[
\begin{array}{ccl}
S'_{<2c} \!\!\! &\approx& \!\!\! \displaystyle 4\lambda \mu^\eta\int_{r_0}^\infty\! x^{-2\eta} e^{\mu\left(c+x\right)} \Big(\Gamma\!\left(1\!-\!\eta,\mu\left(c\!+\!x\right)\right) - \\ & & \displaystyle \Gamma\!\left(1\!-\!\eta,\mu\left(2c\!+\!x\right)\right)  \Big) {\rm d}x.
\end{array}
\]

In order to approximate the above integral, we first expand the integrand for $\mu\left(c\!+\!x\right)\!\rightarrow\!\infty$. Due to the fact that $\mu\!\geq\!\lambda$ and $x\!\geq\!r_0$, the expansion is valid for $\lambda r_0\!\gg\! 1$, which is associated with a large number (on average) of vehicles inside the cell.  
\[
\begin{array}{ccl}
S'_{<2c} \!\!\!&\approx&\!\!\! \displaystyle  \frac{4\lambda e^{-c\mu}r_0^{-3\eta} \left(\eta\!-\!c\mu\!+\!c\eta\mu\!+\!e^{c\mu}\left(c\mu\!-\!\eta\right)\right)}{3\eta\mu} \times \\ & & \displaystyle   \,\,\,\,\,\, {}_2F_1\!\left(3\eta,\eta\!+\!1,3\eta\!+\!1,-\frac{c}{r_0}\right) + \\ & & \displaystyle  \frac{4\lambda r_0^{1-3\eta} \left(1\!-\!e^{-c\mu}\right){}_2F_1\!\left(3\eta\!-\!1,\eta\!+\!1,3\eta,-\frac{c}{r_0}\right)}{\left(3\eta-1\right)}. 
\end{array}
\]

After substituting $\mu\!=\!\frac{\lambda}{1\!-\!\lambda c}$, the above expression can be further approximated for $\lambda c\!\rightarrow\! 0$ and $\frac{c}{r_0}\!\rightarrow\! 0$. Keeping only the dominant term with respect to $r_0$, we end up with 
\[
S'_{<2c} \approx \frac{4\lambda^2 c \, r_0^{1-3\eta}}{3\eta-1} + \frac{2\lambda^3 c^2 r_0^{1-3\eta}}{3\eta-1}.
\]

\noindent 
After summing up $S'_{>2c}$ with $S'_{<2c}$ and scaling the result by six, see equation~\eqref{eq:ThirdMomInterf}, we have 
\begin{equation}
\label{eq:SprimFinal}
6 S'\approx \frac{24\lambda^2 r_0^{2-3\eta}}{\left(2\eta\!-\!1\right)\left(\eta\!-\!1\right)} - \frac{24  \lambda^2 c \, r_0^{1-3\eta}}{3\eta-1} + \frac{12\lambda^3 c^2 r_0^{1-3\eta}}{3\eta-1}. 
\end{equation}

The calculation of $S''\!=\!\int\!\! x^{-\eta}y^{-\eta}z^{-\eta} \rho^{(3)}\!\left(x,y,z\right) {\rm d}x{\rm d}y{\rm d}z$ is more tedious than the calculation of $S'$ because it involves triples instead of pairs of users. It is shown in~\cite{Koufos2018} that the contribution to $S''$ can be split  into two parts: (i) $S_1''$ with the vehicles $x,y,z$ located at the same side with respect to the cell, and (ii) $S_2''$ with the location of one vehicle  being uncorrelated to the locations of the other two because it is located at the opposite side of the cell. $S''\!=\!S_1''\!+\!S_2''$. 

Using the common approximation for the \ac{PCF} and assuming the order $r_0\!<\!x\!<\!y\!<\!z$ for the three vehicles, the contribution to $S_1''$ can be divided into four terms describing the possible separation of distances between each pair of vehicles $\left\{x,y\right\}$ and $\left\{y,z\right\}$.
\[
\begin{array}{ccl}
S_1'' \!\!\!&\approx&\!\!\! \displaystyle \underbrace{12\lambda^2\mu \!\! \int_{r_0}^\infty\!\!\!\int_{x+2c}^\infty\!\int_{y+c}^{y+2c}\!\!\frac{z^{-\eta}y^{-\eta}x^{-\eta}}{e^{\mu\left(z-y-c\right)}}{\rm d}z {\rm d}y {\rm d}x}_{S''_{11}} + \\ & & \displaystyle  \underbrace{12\lambda^3 \!\! \int_{r_0}^\infty\!\!\!\int_{x+2c}^\infty\!\int_{y+2c}^\infty\!\!\!\!\!\! z^{-\eta}y^{-\eta}x^{-\eta} {\rm d}z {\rm d}y {\rm d}x}_{S''_{12}} \, + \\ \!\!\!& &\!\!\! \displaystyle 12\lambda\mu^2 \!\! \int_{r_0}^\infty\!\!\!\int_{x+c}^{x+2c}\!\!\!\!\int_{y+c}^{y+2c}\!\!\frac{z^{-\eta}y^{-\eta}x^{-\eta}}{e^{\mu\left(z-x-2c\right)}}{\rm d}z {\rm d}y {\rm d}x \,\,\, + \\ & & \displaystyle 12\lambda^2\mu \!\!\int_{r_0}^\infty\!\!\!\int_{x+c}^{x+2c}\!\!\!\!\int_{y+2c}^\infty\!\!\frac{z^{-\eta}y^{-\eta}x^{-\eta}}{e^{\mu\left(y-x-c\right)}}{\rm d}z {\rm d}y {\rm d}x, 
\end{array}
\]
where the factor $12\!=\!6\times 2$ accounts for the six different orderings of $x,y,z$ and the factor two is added to account for the three users being at the negative half-axis. 

\noindent 
The term $S''_{11}$ can be approximated as follows
\[
\begin{array}{ccl}
S_{11}'' \!\!\!\!\!\!\!&=&\!\!\!\!\!\!\! \displaystyle 12 \lambda^2\mu \!\! \int_{r_0}^\infty\!\!\!\int_{x+2c}^\infty\!\! x^{-\eta}y^{-\eta} e^w \mu^{\eta-1} \Big(\Gamma\!\left(1\!-\!\eta,w\right) - \\ & & \displaystyle \Gamma\!\left(1\!-\!\eta,w\!+\!c\mu\right) {\rm d}y {\rm d}x\Big) \\ \!\!\!\!\!\!\!&\stackrel{(a)}{\approx}&\!\!\!\!\!\!\! \displaystyle 12\lambda^2 \!\!\!\int\limits_{r_0}^\infty\!\! \Bigg(  \frac{\left(\eta\!-\!c\mu\!+\!\eta c\mu\!+\!e^{c\mu}\!\left(c\mu\!-\!\eta\right)\right) {}_2F_1\!\!\left(\!2\eta,\!1\!+\!\eta,\!2\eta\!+\!1,\!\frac{-c}{2c+x}\!\right)}{2\mu\, \eta\,  e^{c\mu} \, \left(x\!+\!2c\right)^{2\eta}}  \\ & & \displaystyle + \frac{\left(1\!-\!e^{-c\mu}\right)\left(x\!+\!2c\right)^{\!1\!-\!2\eta} {}_2F_1\!\left(\eta\!+\!1,\!2\eta\!-\!1,\!2\eta,\!\frac{-c}{2c+x}\right)}{2\eta-1}\Bigg) {\rm d}x \\ \!\!\!\!\!& &\!\!\!\!\! \displaystyle \stackrel{(b)}{\approx} \frac{12\lambda^3 c\,  r_0^{2-3\eta}}{\left(2\eta-1\right)\left(3\eta-2\right)}, 
\end{array}
\]
where $w\!=\!\mu\left(c\!+\!y\right)$, $(a)$ follows from expanding at  $w\!\rightarrow\!\infty$ before integrating in terms of $y$, and $(b)$ from expanding around $\frac{c}{x}\!\rightarrow\! 0$ before integrating in terms of $x$, then substituting $\mu\!=\!\frac{\lambda}{1-\lambda c}$ and expanding at $\lambda c\!\rightarrow\! 0$. 

The term $S''_{12}$ does not involve any exponential but still, it cannot be expressed in semi-closed form, unless approximations are made in the integrand.
\[
\begin{array}{ccl}
S''_{12} \!\!\!\!\!\!\!& = &\!\!\!\!\!\!\! \displaystyle  12\lambda^3 \!\! \int_{r_0}^\infty\!\!\!\int_{x+2c}^\infty \! \frac{x^{-\eta} y^{-\eta} \left(2c\!+\!y\right)^{1-\eta}}{\eta-1} {\rm d}y {\rm d}x \\  \!\!\!\!\!\!\!& = &\!\!\!\!\!\!\! \displaystyle 
12\lambda^3 \!\!\! \int\limits_{r_0}^\infty \!\!  \frac{x^{\!-\eta} \! \left(2c\!+\!x\right)^{\!1\!-\!2\eta}}{2\left(\eta\!-\!1\right)} \! \Bigg(\! \frac{\left(2c\!+\!x\right) {}_2F_1\!\!\left(\!2\eta\!-\!2,\!\eta,\!2\eta\!-\!1,\!\frac{-2c}{2c+x}\!\right)}{\eta\!-\!1}  + \\  & & \displaystyle \frac{4c}{2\eta-1} \, {}_2F_1\!\left(\!2\eta\!-\!1,\!\eta,\!2\eta,\!\frac{-2c}{2c+x}\!\right) \!\! \Bigg) {\rm d}x \\ \!\!\!\!\!\!\!&\approx&\!\!\!\!\!\!\! \displaystyle \frac{2\lambda^3 r_0^{3-3\eta}}{\left(\eta-1\right)^3} - \frac{24\lambda^3 c\, r_0^{2-3\eta}}{\left(\eta\!-\!1\right)\left(2\eta\!-\!1\right)} \!+\! \frac{12\lambda^3 c^2 r_0^{1-3\eta}\left(9\eta\!-\!7\right)}{\left(\eta\!-\!1\right)\left(3\eta\!-\!1\right)},
\end{array}
\]
where the approximation is due to expansion $\frac{c}{x}\!\rightarrow\! 0$ before integrating. 

\noindent 
The other two terms of $S''_1$ can be approximated similarly, 
\[
\begin{array}{ccl}
S''_{13} \!\!\!&\approx&\!\!\!  \displaystyle \frac{12 \lambda^3 c^2 r_0^{1-3\eta}}{3\eta-1} \\ 
S''_{14} \!\!\!&\approx&\!\!\!  \displaystyle \frac{12 \lambda^3 c r_0^{2-3\eta}}{\left(3\eta-2\right)\left(\eta-1\right)} - \frac{24 \lambda^3 c^2\eta r_0^{1-3\eta}}{\left(3\eta-1\right)\left(\eta-1\right)}. 
\end{array}
\]

After adding up the approximations for the four terms consisting $S_1''$, we end up with  
\begin{equation}
\label{eq:S1prim2}
\begin{array}{ccl}
S''_1 \!\!\!\!\!&\approx&\!\!\!\!\! \displaystyle \frac{2\lambda^3 r_0^{3-3\eta}}{\left(\eta-1\right)^3} \!-\! \frac{12\lambda^3 c\, r_0^{2-3\eta}}{\left(\eta\!-\!1\right)\left(2\eta\!-\!1\right)} \!+\! \frac{96\lambda^3 c^2 r_0^{1-3\eta}}{3\eta-1}. 
\end{array}
\end{equation}

Assuming $x\!<\!y\!<\!z$ and the user $x$ at the opposite side of the cell as compared to the users $y,z$, the term $S_2''$ is
\[
\begin{array}{ccl}
S_2'' \!\!\!&\approx&\!\!\! \displaystyle 12\lambda\int_{-\infty}^{-r_0} \!\! \left|x\right|^{-\eta}{\rm d}x \,  \bigg(\lambda^2\!\!\int_{r_0}^\infty\!\!\int_{y+2c}^\infty \!\!\!\! y^{-\eta} z^{-\eta} {\rm dz} {\rm dy} \, + \\ & & \displaystyle \lambda\mu\!\! \int_{r_0}^\infty\!\!\int_{y+c}^{y+2c}\!\! \frac{y^{-\eta} z^{-\eta}}{e^{\mu\left(z-y-c\right)}} {\rm d}z {\rm d}y \bigg), 
\end{array}
\]
where the factor $12\!=\!6\times 2$ is due to six different orderings of vehicles and the scaling by two is used to describe the case where the sides, with respect to the cell, of the user $x$ and of the pair $\left\{y,z\right\}$ are inter-changed. 

The two integrals inside the parenthesis can be approximated similarly to the term $S'$. After integrating the first, and approximating the second for $\mu\left(x\!+\!c\right)\!\rightarrow\!\infty$ and $\lambda c\!\rightarrow\! 0$ we have 
\[
\begin{array}{ccl}
S_2'' \!\!\!\!\!&\approx&\!\!\!\!\! \displaystyle \frac{12\lambda r_0^{1-\eta}}{\eta-1} \Bigg(\frac{\lambda^2 r_0^{1-2\eta}}{2\left(\eta\!-\!1\right)}\bigg( \frac{r_0{}_2F_1\!\left(2\left(\eta\!-\!1\right),\eta,2\eta\!-\!1,-2b\right)}{\eta-1} + \\  \!\!\!\!\!\!\!\!\!\!\!& &\!\!\!\!\!\!\!\!\!\!\! \displaystyle \frac{4c{}_2F_1\!\!\left(\!2\eta\!-\!1,\!\eta,\!2\eta,\!-2b\!\right)}{2\eta-1} \bigg) \!\!+\!\!  \frac{\lambda^2 c  r_0^{\!1-2\eta}{}_2F_1\!\!\left(\!\eta\!+\!1\!,2\eta\!-\!1,\!2\eta,\!-b\!\right)}{2\eta-1} \\ \!\!\!\!\!& &\!\!\!\!\! \displaystyle + \frac{\lambda^2c^2\, r_0^{-2\eta}}{2} \bigg( \frac{\lambda r_0 \,  {}_2F_1\!\left(2\eta\!-\!1,\!1\!+\!\eta,\!2\eta,\!-b\right)}{2\eta-1} \, - \\ & & \displaystyle  \frac{\left(\eta\!-\!2\right) {}_2F_1\!\left(2\eta,\!1\!+\!\eta,\!2\eta\!+\!1,\!-b\right)}{2\eta}\bigg) \Bigg). 
\end{array}
\]

\noindent 
Expanding the above expression for small $b\!=\! \frac{c}{r_0}$ yields 
\begin{equation}
\label{eq:S2prim}
\begin{array}{ccl}
S_2'' \!\!\!&\approx&\!\!\! \displaystyle \frac{6\lambda^3 r_0^{3-3\eta}}{\left(\eta-1\right)^3} -\frac{12 \lambda^3 c \, r_0^{2-3\eta}}{\left(2\eta-1\right)\left(\eta-1\right)}. 
\end{array}
\end{equation}

Now, we can express the term $S''$ equal to the sum of~\eqref{eq:S1prim2} and~\eqref{eq:S2prim}
\begin{equation}
\label{eq:Sprim2}
\begin{array}{ccl}
S'' \!\!\!\!\!&\approx&\!\!\!\!\! \displaystyle \frac{8\lambda^3 r_0^{3-3\eta}}{\left(\eta-1\right)^3} -\frac{24 \lambda^3 c \, r_0^{2-3\eta}}{\left(2\eta\!-\!1\right)\left(\eta\!-\!1\right)} \!+\! \frac{96\lambda^3 c^2 r_0^{1-3\eta}}{3\eta-1}. 
\end{array}
\end{equation}

Substituting~\eqref{eq:SprimFinal} and~\eqref{eq:Sprim2} into~\eqref{eq:ThirdMomInterf}, and doing some rearrangement allows us to  approximate the third moment of interference as 
\[
\begin{array}{ccl}
\mathbb{E}\!\left\{\mathcal{I}^3\right\} \!\!\!\!\! &\approx& \!\!\!\!\! \displaystyle  \frac{12\lambda r_0^{1-3\eta}}{3\eta-1} \left(1\!-\!2\lambda c \!+\! 9\lambda^2 c^2\right) + \\ & & \displaystyle \frac{24\lambda^2 r_0^{2-3\eta}}{\left(2\eta\!-\!1\right)\left(\eta\!-\!1\right)}\left(1-\lambda c\right) +  \frac{8\lambda^3 r_0^{3-3\eta}}{\left(\eta\!-\!1\right)^3}. 
\end{array}
\]

Using the approximation for the variance, $\mathbb{V}\!\left\{\mathcal{I}\right\}\!=\! \frac{4\lambda r_0^{1-2\eta}}{2\eta-1} \left(1-\lambda c+\frac{1}{2}\lambda^2c^2\right)$, see~\cite{Koufos2018}, the third central moment,  $\mathbb{E}\!\left\{\mathcal{I}^3_c\right\}\!=\!\mathbb{E}\left\{\left(\mathcal{I}-\mathbb{E}\!\left\{\mathcal{I}\right\}\right)^3\right\}$, becomes 
\[ 
\mathbb{E}\!\left\{\mathcal{I}_c^3\right\} \approx  \displaystyle  \frac{12\lambda r_0^{1-3\eta}}{3\eta-1} \left(1\!-\!2\lambda c \!+\! 9\lambda^2 c^2\right) - \frac{12 \lambda^4 c^2 r_0^{2-3\eta}}{\left(2\eta-1\right)\left(\eta-1\right)}.
\]

Using the approximations for the third central moment and the variance above, the skewness is 
\[
\begin{array}{ccl}
\mathbb{S}\!\left\{\mathcal{I}\right\} \!\!\!\!\!&\approx&\!\!\!\!\! \displaystyle \left(\frac{12\lambda r_0^{1-3\eta}\left(1\!-\!2\lambda c \!+\! 9\lambda^2 c^2\right)}{3\eta-1} - \frac{12 \lambda^4 c^2 r_0^{2-3\eta}}{\left(2\eta\!-\!1\right)\left(\eta\!-\!1\right)} \right) \times \\ & &  \displaystyle \left(\frac{4\lambda r_0^{1-2\eta}}{2\eta-1} \left(1-\lambda c+\frac{1}{2}\lambda^2c^2\right) \right)^{-\frac{3}{2}}. 
\end{array}
\]

\noindent 
Expanding the above expression for $\lambda c\!\rightarrow\! 0$ yields 
\begin{equation}
\label{eq:SkewnessApp}
\mathbb{S}\!\left\{\mathcal{I}\right\} \approx \frac{12 \lambda r_0^{1-3\eta}}{3\eta-1} \left( \frac{4\lambda r_0^{1-2\eta}}{2\eta-1} \right)^{-\frac{3}{2}} \left(1 - \frac{\lambda c}{2} \right),
\end{equation}
where $\left(1-\frac{\lambda c}{2}\right)$ is the correction as compared to the skewness of interference due to a \ac{PPP} of intensity $\lambda$.

\end{document}